 \documentclass[sigconf]{acmart}
\AtBeginDocument{%
  \providecommand\BibTeX{{%
    \normalfont B\kern-0.5em{\scshape i\kern-0.25em b}\kern-0.8em\TeX}}}

\copyrightyear{2024}
\acmYear{2024}
\setcopyright{rightsretained}
\acmConference[IUI '24]{29th International Conference on Intelligent User Interfaces}{March 18--21, 2024}{Greenville, SC, USA}
\acmBooktitle{29th International Conference on Intelligent User Interfaces (IUI '24), March 18--21, 2024, Greenville, SC, USA}
\acmDOI{10.1145/3640543.3645148}
\acmISBN{979-8-4007-0508-3/24/03}

\usepackage{multirow}
\usepackage{framed,enumitem} 
\usepackage{graphicx}
\usepackage{xspace}
\usepackage{makecell}
\usepackage{siunitx}
\usepackage{subfigure}
\usepackage{stfloats}
\newcommand{\intent}[1]{$\mathbf{D_{intent}}$}
\newcommand{\depth}[1]{$\mathbf{D_{depth}}$}
\newcommand{\acc}[1]{$\mathbf{D_{acc}}$}
\newcommand{\trans}[1]{$\mathbf{D_{trans}}$}
\newcommand{\refuse}[1]{$\mathbf{D_{refuse}}$}
\newcommand{\ethic}[1]{$\mathbf{D_{ethic}}$}
\newcommand{\format}[1]{$\mathbf{D_{format}}$}

\newcommand{\Trepeat}[1]{$\mathbf{T_{repeat}}$}
\newcommand{\Tspecify}[1]{$\mathbf{T_{specify}}$}
\newcommand{\Terror}[1]{$\mathbf{T_{error}}$}
\newcommand{\Tadapt}[1]{$\mathbf{T_{adapt}}$}
\newcommand{\Tno}[1]{\textbf{No Tactic}}

\newcommand{\kw}[1]{Kruskal-Wallis}

\begin{document}

\title[Understanding Users' Dissatisfaction with ChatGPT Responses]{Understanding Users’ Dissatisfaction with ChatGPT Responses: Types, Resolving Tactics, and the Effect of Knowledge Level}


\author{Yoonsu Kim}
\email{yoonsu16@kaist.ac.kr}
\affiliation{%
  \institution{Graduate School of AI, KAIST}
  \city{Daejeon}
  \country{Republic of Korea}
}

\author{Jueon Lee}
\email{audreylee@snu.ac.kr}
\affiliation{%
  \institution{College of Liberal Studies, SNU}
  \city{Seoul}
  \country{Republic of Korea}
}
\author{Seoyoung Kim}
\email{youthskim@kaist.ac.kr}
\affiliation{%
  \institution{School of Computing, KAIST}
  \city{Daejeon}
  \country{Republic of Korea}
}
\author{Jaehyuk Park}
\email{jp@kdischool.ac.kr}
\affiliation{%
  \institution{School of Public Policy and Management, KDI}
  \city{Sejong}
  \country{Republic of Korea}
}

\author{Juho Kim}
\email{juhokim@kaist.ac.kr}
\affiliation{%
  \institution{School of Computing, KAIST}
  \city{Daejeon}
  \country{Republic of Korea}
}

\renewcommand{\shortauthors}{Yoonsu Kim et al.}





\begin{abstract}
Large language models (LLMs) with chat-based capabilities, such as ChatGPT, are widely used in various workflows. However, due to a limited understanding of these large-scale models, users struggle to use this technology and experience different kinds of dissatisfaction. 
Researchers have introduced several methods, such as prompt engineering, to improve model responses.
However, they focus on enhancing the model's performance in specific tasks, and little has been investigated on how to deal with the user dissatisfaction resulting from the model's responses. 
Therefore, with ChatGPT as the case study, we examine users' dissatisfaction along with their strategies to address the dissatisfaction.
After organizing users' dissatisfaction with LLM into seven categories based on a literature review, we collected 511 instances of dissatisfactory ChatGPT responses from 107 users and their detailed recollections of dissatisfactory experiences, which we released as a publicly accessible dataset. 
Our analysis reveals that users most frequently experience dissatisfaction when ChatGPT fails to grasp their intentions, while they rate the severity of dissatisfaction related to accuracy the highest. 
We also identified four tactics users employ to address their dissatisfaction and their effectiveness.
We found that users often do not use any tactics to address their dissatisfaction, and even when using tactics, 72\% of dissatisfaction remained unresolved.
Moreover, we found that users with low knowledge of LLMs tend to face more dissatisfaction on accuracy while they often put minimal effort in addressing dissatisfaction.
Based on these findings, we propose design implications for minimizing user dissatisfaction and enhancing the usability of chat-based LLM.
\end{abstract}

\maketitle


\section{Introduction}
\begin{figure*}[!ht]
    \centering
    \includegraphics[width=1.0\textwidth]{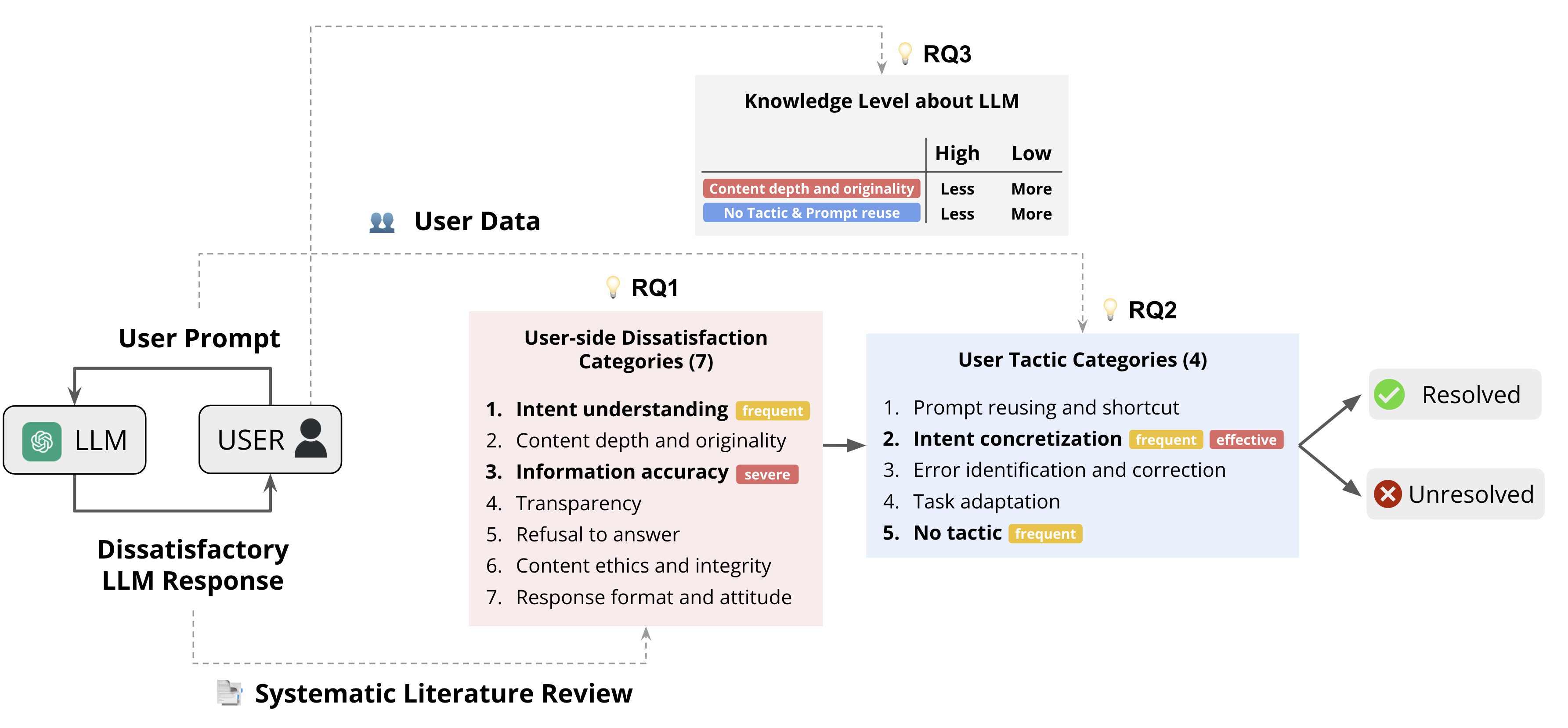}
    \caption{Overview of our research questions and findings.}
    \Description{This figure outlines our research question and findings. It contains seven components: User, LLM, User-side dissatisfaction category (RQ1), User Tactic Category (RQ2), Knowledge Level about LLM (RQ3), Solved, and Unsolved. User experiences dissatisfaction from LLM responses, and uses prompts to try to solve this dissatisfaction. Through systematic literature review, we identified seven user-side dissatisfaction categories stemming from LLM responses. The specific categories are explained in Table 1. Through user data of ChatGPT experiences, we found that the dissatisfaction category for 'intent understanding' was the most frequent, while the 'information accuracy' category caused most severe dissatisfaction(RQ1). Five user tactic categories were found in our qualitative analysis of user data, and the tactic 'intent concretization' was found to be both frequent and effective(RQ2). The specific tactic categories are explained in Table 3. User may or may not solve their dissatisfaction with their tactics. Depending on how knowledgeable the users are about the mechanism of LLMs, users with less knowledge were more likely to experience dissatisfaction related to content depth and originality, while also frequently employing no tactics at all or reusing prompts (RQ3).}
    \label{fig:overview}
\end{figure*}
Large Language Models (LLM) have exhibited remarkable performance across various tasks (e.g., language generation~\cite{openai2023gpt4} and reasoning ~\cite{huang2022towards}), and they have become more accessible with integration into chat interfaces and instruction tuning~\cite{ouyang2022training}, such as ChatGPT~\footnote{https://chat.openai.com/}.
As a result, many people are increasingly incorporating this technology into their workflows across various domains such as education ~\cite{kasneci2023chatgpt, sallam2023chatgpt}, healthcare ~\cite{sallam2023chatgpt, kumar2023analysis, mann2023artificial}, and law ~\cite{blair2023can, nay2022law}.

When using a chat-based LLM, natural language prompts play a crucial role because they are the primary medium for interaction between the user and the model ~\cite{zhou2022large, dang2022prompt, zamfirescu2023johnny}. 
Accordingly, prompt engineering---aimed at enhancing the quality of model responses to get desired responses from the model---has been a popular stream of research. 
As various people use LLMs in their workflows, researchers and practitioners have published various guidelines, tools, books, and even online courses for prompt engineering, not only for developers but also for laypeople~\cite{zhou2022large, reynolds2021prompt, white2023prompt, chatGPTM36online}.

However, despite the proliferation of these resources, end-users often encounter dissatisfaction during conversations with LLMs.
When end-users have limited knowledge about LLMs, they may have incorrect expectations about the model's behavior, which can further contribute to their dissatisfaction.
This dissatisfaction may arise from various known limitations of LLMs, including hallucination ~\cite{bang2023multitask, ji2023survey, liu2023trustworthy}, inconsistency ~\cite{jang2023consistency, elazar2021measuring, liu2023trustworthy}, unfavorable tone and format~\cite{ray2023chatgpt, azaria:hal-03913837, yang2023exploring}, and lack of transparency ~\cite{thirunavukarasu2023large, bubeck2023sparks}.
In addition, such dissatisfaction can become more critical when end-users utilize LLMs for practical purposes.

Little previous research, however, has investigated users' dissatisfaction during conversations with LLMs.
In particular, existing prompt engineering techniques mainly focus on enhancing the model's performance in specific tasks, and little has been investigated on how users should respond to dissatisfactions they face from LLMs' responses during the conversation.
Therefore, in our research, with ChatGPT as the case study, we aim to understand the dissatisfaction experienced by the users during the conversations. 
We focus on situations where users seek practical assistance from ChatGPT within their workflows (e.g., translation, email writing, and programming) rather than situations where users intentionally provoke dissatisfactory responses from ChatGPT and test its boundaries and limitations.
Specifically, we explore the types of dissatisfaction users experience during the conversation, how serious each type of dissatisfaction is, and how users address dissatisfaction in the subsequent prompts. 
Furthermore, building upon prior research that demonstrated how users' experiences with technological failure depend on their knowledge of that technology in the context of conversational agent ~\cite{luger2016like}, we investigate how dissatisfaction and user responses vary based on the user's knowledge level of LLMs.

At first, we conducted a systematic literature review of papers dealing with limitations and challenges associated with LLMs and identified seven user-side dissatisfaction categories stemming from LLM responses (Table ~\ref{tab:slr-result}).
Then, using ChatGPT as a case study, we collected how much users confront these seven dissatisfaction categories and how they respond to them during actual conversations through our data collection system (Figure ~\ref{fig:system}).
As a result, we collected 307 ChatGPT conversation logs from 107 respondents, which contained 511 user-side dissatisfactions on ChatGPT responses.
Through a quantitative analysis, we found that users most frequently experienced dissatisfaction in terms of ChatGPT's poor understanding of users' intent, while users felt the most severe dissatisfaction related to inaccuracies in information. 
We also conducted a qualitative analysis of users' behavior to address the dissatisfaction at subsequent prompts, which resulted in four tactic categories (Table ~\ref{tab:tactic-result}): `prompt reusing', `intent concretization', `error identification and correction', `task adaptation', and `no tactic'.
Moreover, we analyzed differences in dissatisfaction and tactics across the users' knowledge levels on LLMs and confirmed that low-knowledge users more frequently experienced dissatisfaction regarding ChatGPT's responses being too general and lacking originality. 
We also observed that low-knowledge users often resorted to `no tactic' or `prompt reusing', which involved minimal efforts in prompt crafting when they experienced dissatisfaction.

Based on our findings, we suggest design implications to improve the usability of LLMs for users, leveraging the occurrence of dissatisfaction and corresponding tactics during the conversation. 
We also suggest that the responses of LLMs could be more tailored to the user's knowledge level.
Furthermore, we release the actual user data we collected as a publicly available dataset~\footnote{https://chatgpt-analysis.kixlab.org} to aid relevant research.
The contributions of our research are as follows: 
\begin{itemize}
    \item Categorization and analysis of user-side dissatisfaction and corresponding tactics at the conversational turn level.
    \item Investigation of how dissatisfaction and tactics appear differently depending on users’ knowledge level of LLMs.
    \item A dataset containing specific user interactions and their experiences of dissatisfaction in actual conversations with ChatGPT, thereby offering resources for further research on user-centric LLMs.
\end{itemize}
\section{Related Work}
We review related work in (1) limitations and user challenges in LLMs and (2) user's strategies to overcome those challenges in Language Models. 

\subsection{Limitations and User Challenges in LLMs}
A rich body of previous work has addressed various limitations associated with language models, including hallucination ~\cite{bang2023multitask, ji2023survey, liu2023trustworthy}, inconsistency in reasoning~\cite{jang2023consistency, elazar2021measuring, liu2023trustworthy}, and numerical computation~\cite{qian2022limitations, yuan2023well}.
Zhao et al. ~\cite{zhao2023survey} reviewed major challenges in recent large language models in terms of three basic types of ability of LLMs: language generation, knowledge utilization, and complex reasoning.
Borji ~\cite{borji2023categorical} organized ChatGPT's failures into eleven distinct categories, including reasoning, factual errors, math, coding, and bias.

However, how users actually experience may be different from LLM's failures. Thus, several studies investigated challenges that can be experienced from the user's side ~\cite{borji2023categorical,sallam2023chatgpt, behrooz2023hci}.
Behrooz ~\cite{behrooz2023hci} points out the core challenges of research chatbots like OpenAI’s ChatGPT, Meta AI’s BlenderBot, and Google’s LaMDA, especially related to user perceptions.
These challenges encompass the lack of conversational context ~\cite{brown2015people, van2007comments}, the speaker perception void ~\cite{holtgraves2008language}, and the lack of expectation baseline ~\cite{skjuve2019help}.

While a stream of research has explored the limitations of language models and the user challenges when interacting with them, there is a lack of comprehensive categorization of the user-side dissatisfaction and how often and seriously users experience each dissatisfaction in the context of users' actual conversation situations. 
Understanding the user-side dissatisfactions arising from practical usage can provide insights into building LLMs with better usability. To this end, our paper investigates how users experience dissatisfaction and the severity of these dissatisfactions by analyzing users' conversation logs with LLMs. 

\subsection{User's Strategies to Overcome Challenges in Language Models}
To improve the usability of language models, it is important to understand users' current practices to overcome the challenges they face. For this, previous research has delved into how users react and overcome challenges encountered while interacting with various language models. 
Porcheron et al. ~\cite{porcheron2018voice} and Luger et al. ~\cite{luger2016like} examined how users interact with a conversational agent in voice user interfaces (VUI). 
Specifically, Myers et al. ~\cite{myers2018patterns} identified ten main categories of tactics users employ to overcome challenges encountered in VUI, and discovered patterns of tactics.
Although LLMs and VUIs share the same characteristic in that users communicate with AI agents via natural language, how users overcome challenges may differ as LLMs use text prompting, which may allow more careful prompting strategies compared to VUIs.

Accordingly, prompt engineering techniques have been extensively studied to address challenges in LLMs~\cite{zhou2022large, reynolds2021prompt, white2023prompt, wei2022chain, madaan2023self}.
For instance, Chain-of-Thought Prompting (CoT) is renowned for improving LLM's reasoning performance by integrating intermediate reasoning steps into prompts ~\cite{wei2022chain}.
Building upon the effectiveness of CoT, researchers have explored variants like Zero-shot CoT ~\cite{kojima2022large}, Auto-CoT~\cite{zhang2022automatic}, and Self-Consistency (CoT-SC) ~\cite{wang2022self} and showed that those methods can mitigate LLM's deficiency in reasoning. Specifically, CoT-SC is also known for mitigating LLM's inconsistency issue.
Madaan et al. ~\cite{madaan2022language} also showed that transforming a certain task into a code generation task can be effective in addressing reasoning and inconsistency issues in LLMs.

However, previous studies investigate how to enhance the model's performance in specific tasks (e.g., reasoning), and they rarely addressing how to handle the dissatisfaction experienced by users during conversations with LLM, stemming from the responses they receive.
Therefore, in this paper, we investigate users' behaviors when they encounter dissatisfaction from their actual conversations with LLM. Through this, we analyze users' tactics to address their dissatisfaction and their effectiveness. This will provide insights into how LLM and its interface can be further developed to aid users when they encounter dissatisfaction in the middle of the conversation.
\section{Systematic Literature Review: Categorizing User-side Dissatisfaction} \label{sec:slr}

To understand and categorize the dissatisfaction points that users encounter when using LLMs for practical purposes, we conducted a systematic literature review to investigate the challenges, limitations, and failures identified in previous research within the LLM context. 
We focused on user-side dissatisfaction experiences directly arising from LLM responses. 
For this purpose, we scrutinized a total of 59 papers and conducted qualitative coding, which resulted in 19 codes representing user-side dissatisfaction points from LLM responses. These points were subsequently categorized into seven themes (Table~\ref{tab:slr-result}). 
The seven themes were provided as multiple-choice items in our data collection, allowing users to select the dissatisfaction points they have experienced from LLM responses. 

\subsection{Search Keywords}\label{sec:search-process}
We first conducted an extensive search on Google Scholar~\footnote{https://scholar.google.com}, ACM Digital Library~\footnote{https://dl.acm.org}, and arXiv~\footnote{https://arxiv.org} using the combination of ``Large Language Models(LLMs),'' and ``ChatGPT,'' with ``Challenges,'' ``Limitations,'' and ``Difficulties'' as search keywords.
The reason we specifically included ChatGPT as a search keyword is because ChatGPT has been one of the most extensively used LLMs and has been widely adopted across a variety of domains, such as the medical domain and education. 
Considering the temporal progress in LLM technologies, we restricted the search period to after 2021.
To not exclude papers that might be relevant but do not explicitly contain our search keywords, we extended our search by traversing the citation graph of the initial set of papers. 
We explored the papers that are either cited by or cite the papers within our initial set and gathered any papers that discuss user-side dissatisfaction, challenges, or difficulties with the use of LLMs, as well as instances of LLM failures.

\subsection{Exclusion Criteria and Filtering}
As the result of the search process (Section~\ref{sec:search-process}), we collected 1,249 papers. After removing duplicates, we had 866 papers. To focus on user-side dissatisfaction with LLM responses, we set the 4 exclusion criteria and filtered papers based on them.
\begin{enumerate}
    \item[\textbf{EC1.}] We excluded papers that used the terms "limitation," "challenge," or "difficulty" in a general sense, not specifically about LLMs.
    \item[\textbf{EC2.}] We excluded papers that focused solely on the technical challenges or limitations of LLMs.
    \item[\textbf{EC3.}] We excluded papers that discussed potential risks of LLM usage, such as the overreliance of students on LLMs for learning ~\cite{Rahman2023ChatGPTFE, Joshi2023ChatGPTIT} or the potential for privacy issues~\cite{Gupta2023FromCT}.
    \item[\textbf{EC4.}] We excluded papers that discuss the difficulty of tuning or maintaining LLMs that are not directly related to LLMs responses.
\end{enumerate}
We filtered papers following these criteria, resulting in 59 papers.
This allows us to include papers that discuss the practical application of LLMs in specific domains or workflows intended to enhance productivity, which resulted in diverse fields such as education, healthcare, and research. 
\begin{table*}[!ht]
\tiny{
\begin{center}
\resizebox{\textwidth}{!}{%
\def\arraystretch{1.5}%
\begin{tabular}{p{0.15\columnwidth}p{0.25\columnwidth}p{0.55\columnwidth}c}
\toprule
    \textbf{Category (7)} & \textbf{Description} & \textbf{Code (19)} & \textbf{Example} \\ 
\midrule
    \multirow{3}{0.15\columnwidth}{\makecell[l]{Intent \\ Understanding \\ \textbf{(\intent{})}}} & \multirow{3}{0.25\columnwidth}{This response does not correctly reflect the user’s intent, instruction, or context.} & C1. Response does not meet users' intent or instruction. & \cite{kaddour2023challenges} \\ 
\cline{3-4}  
&  & C2. Response is not aligned with the user's context. &  \cite{rao2023evaluating} \\ 
\cline{3-4} 
 &  & C17. The tone or communication style is disappointing. & \cite{ray2023chatgpt} \\ 
\hline{}
 
\multirow{3}{0.15\columnwidth}{\makecell[l]{Content Depth \\ and Originality \\ \textbf{(\depth{})}}} & \multirow{3}{0.25\columnwidth}{This response is overly general, lacks originality, or needs more diversity.} & C3. Response is too general. & \cite{borji2023categorical} \\ 
 &  & C4. Response lacks originality. & \cite{kitamura2023chatgpt} \\ 
 \cline{3-4} 
 &  & C5. Response lacks information. & \cite{kumar2023analysis} \\ 
 \hline{}

\multirow{6}{0.15\columnwidth}{\makecell[l]{Information \\ Accuracy \\ \textbf{(\acc{})}}} & \multirow{6}{0.25\columnwidth}{This response contains false/inaccurate information or inconsistency.} & C6. The response contains incorrect information. & \cite{azaria:hal-03913837} \\ 
\cline{3-4} 
&  & C7. Response is based on training data cut off at a certain date, and has limited access to newly created data. & \cite{yeo2023assessing} \\ 
 \cline{3-4}
 &  & C8. Response is inconsistent. & \cite{alkhamissi2022review} \\ 
 \cline{3-4}
 &  & C9. ChatGPT struggles with reasoning. & \cite{zhang2023small} \\ 
 \cline{3-4} 
 &  &  C10. (Hallucination) ChatGPT fabricates contents that conflict with the source content or cannot be verified from existing sources. & \cite{ji2023survey} \\ 
 \cline{3-4} 
 &  & C19. (Sycophancy) ChatGPT excessively conforms to the user. & \cite{perez2022discovering} \\ 
\hline{}

Transparency \textbf{(\trans{})} & It is difficult to understand the underlying reasoning or criteria of this response. & C11. It's difficult to understand the reasons, criteria, logic, and evidence behind the responses. & \cite{thirunavukarasu2023large} \\ 
\hline{}

\multirow{3}{0.15\columnwidth}{\makecell[l]{Refusal to Answer \\ \textbf{(\refuse{})}}} & \multirow{3}{0.25\columnwidth}{ChatGPT avoids answering by saying something similar to ``As a language model, I am not capable …''} & C12. ChatGPT avoids giving its own opinion by saying something similar to ``As a language model, I am not capable …'' & \cite{borji2023categorical} \\ 
\cline{3-4} 
 &  & C13. ChatGPT avoids talking about difficult or controversial issues by saying something similar to ``As a language model, I am not capable ...'' & \cite{bang2023multitask} \\ 
\cline{3-4}
 &  & C7. Response is based on training data cut off at a certain date, and has limited access to newly created data. & \cite{guo2023close} \\
\hline{}
 
\multirow{3}{0.15\columnwidth}{\makecell[l]{Content Ethics \\ and Integrity \\ \textbf{(\ethic{})}}} & \multirow{3}{0.25\columnwidth}{This response contains unlawful, unethical, harmful, or biased content.} & C14. Response contains unlawful content & \cite{liu2023trustworthy} \\ 
\cline{3-4}
&  & C15. Response contains unethical, harmful content. & \cite{weidinger2021ethical} \\ 
\cline{3-4} 
&  & C16. Response contains biased content. & \cite{cao2023comprehensive} \\
\hline{}

\multirow{3}{0.15\columnwidth}{\makecell[l]{Response Format \\ and Attitude \\ \textbf{(\format{})}}} & \multirow{3}{0.25\columnwidth}{The format of this response — including but not limited to tone, length, structure, and attitude — is disappointing.} & C17. The tone or communication style is disappointing. & \cite{guo2023close} \\ 
\cline{3-4} 
&  & C18. Response is overly detailed or too long & \cite{yang2023exploring} \\ 
\cline{3-4} 
&  & C19. (Sycophancy) ChatGPT excessively conforms to the user. & \cite{bang2023multitask} \\ 

\bottomrule
\end{tabular}
}
\end{center}
}
\caption{7 category and corresponding 19 codes of user-side dissatisfaction from LLM Responses.}
\Description{This table summarizes the result of systematic literature review on user dissatisfaction. There are four columns: Category(7), Description, Code(19), and Example. Each dissatisfaction category has one description and one or more specific dissatisfaction codes. Examples are research papers related to each code.}
\label{tab:slr-result}
\end{table*}

\subsection{Analysis Procedure}
To analyze and categorize the user-side dissatisfaction from LLM responses, our initial step involved reading 59 papers and compiling a comprehensive list related to user-side dissatisfaction, challenges, or difficulties with the use of LLMs, as well as instances of LLM failures. 
Two authors then independently conducted open coding on the compiled list. 
Our primary focus was on identifying aspects of user-side dissatisfaction that emanated from interactions with LLM responses. 
Following the individual open coding phase, the two authors engaged in collaborative and iterative discussions. 
These discussions were instrumental in consolidating and refining the initially identified codes. 
The authors worked together to ensure that the codes accurately captured the nuances of user dissatisfaction associated with LLM responses. 
Subsequently, to establish relationships among these codes, all authors participated in axial coding~\cite{strauss1998basics}. 
This involved a series of successive discussions aimed at clustering the individual codes into broader, more abstract categories. 
The goal was to identify common threads and overarching themes that emerged from the data. 
The axial coding process culminated in the consolidation of the identified aspects of user-side dissatisfaction into seven main themes. (Table ~\ref{tab:slr-result}) These themes encapsulated the various dimensions of user dissatisfaction when interacting with LLM responses. 
The dissatisfaction themes were later used when collecting data from users, which is explained in detail in Section ~\ref{sec:data}.

\subsection{Result: Categorizing User-side Dissatisfaction}

We categorized the various aspects of user dissatisfaction arising from LLM responses into 19 distinct codes, further organized into seven overarching themes. The detailed information is denoted in Table~\ref{tab:slr-result}. All paper lists are in the Appendix ~\ref{appendix:slr_all}.

\smallskip
\noindent
\textbf{Theme 1. Intent Understanding (\intent{})}
This theme encompasses issues related to LLM's failure to correctly interpret or reflect the user's intent, instructions, or context. Three codes (C1, C2, C17) fall into this theme. 
LLM outputs often fail to align with the users' needs and expectations \cite{kaddour2023challenges}. ChatGPT has been found to suggest unnecessary out-of-context actions in medical use \cite{rao2023evaluating}, and to use the wrong tone or be excessively literal due to its low understanding of non-literal language such as sarcasm \cite{ray2023chatgpt}.

\smallskip
\noindent
\textbf{Theme 2. Content Depth and Originality (\depth{})}
Users experienced this type of dissatisfaction when they expected more in-depth and creative answers catered to their specific needs, but LLM gave responses that were perceived as overly general, lacking originality, or requiring more diversity. ChatGPT rarely diverges from the topic, generating less diverse content than humans~\cite{borji2023categorical, guo2023close}. 
Concerns rise on unvarying and repetitive ChatGPT outputs which are results of generation based on past data~\cite{kitamura2023chatgpt}.
ChatGPT showed weaknesses in providing practical examples in academic writing \cite{kumar2023analysis}.

\smallskip
\noindent
\textbf{Theme 3. Information Accuracy (\acc{})}
Dissatisfactions related to false, outdated, or inaccurate information in responses fall under this theme. In addition, inconsistencies within one response or in conversation beyond one answer also belong to this theme. Users were dissatisfied when LLMs provided incorrect or conflicting information, eroding trust in the system's reliability.
ChatGPT is incompetent in correctly calculating large numbers ~\cite{azaria:hal-03913837}, and bases its answers on training data up to a certain point in the past - September 2021 is the cutoff in the latest released version of ChatGPT- therefore generating outdated and wrong information when facts change over time~\cite{yeo2023assessing}.
Language models are known to show inconsistency in their claims and explanations~\cite{alkhamissi2022review}. 
ChatGPT has limited reasoning capabilities, including inductive, spatial, and mathematical reasoning~\cite{zhang2023small, bang2023multitask}. Hallucination, the generation of absurd output that contradicts the source or cannot be verified from it, is a threat in real-world applications since the wrong output can cause harm when people trust the outcome of LLMs without further inspection~\cite{ji2023survey}.
Sycophancy, a behavior where LLMs contradict their original output in order to agree with human input, is also a reason for concern about inaccurate and trustworthy generation~\cite{perez2022discovering}.

\smallskip
\noindent
\textbf{Theme 4. Transparency (\trans{})}
Users experiencing difficulties in understanding the underlying reasoning or criteria behind LLM responses led to dissatisfaction related to transparency. Users desired more transparency in how the language model generated its answers, especially when complex or critical information was involved. 
The 'black box' nature of LLMs makes it difficult for users to interpret the reasons behind their outputs.

\smallskip
\noindent
\textbf{Theme 5. Refusal to Answer (\refuse{})}
Responses where LLMs avoided providing answers, often using phrases like ``As a language model, I am not capable...'' or similar, were categorized under this theme. Users were frustrated when the system declined to provide information or guidance.
ChatGPT may refrain from giving its direct opinion ~\cite{borji2023categorical}, and refuse to verify if a claim can be considered misinformation when the claim is closely related to social issues~\cite{bang2023multitask}. Refusing is also found in questions regarding information in a time point outside ChatGPT's training data cutoff~\cite{guo2023close}.

\smallskip
\noindent
\textbf{Theme 6. Content ethics and integrity (\ethic{})}
This theme represents the presence of unlawful, unethical, harmful, or biased content in LLM responses. 
Illegal and dangerous information was found to be accessible through LLMs ~\cite{yang2023exploring}, as well as stereotypes, discriminatory views, and performance disparity in certain groups ~\cite{weidinger2021ethical}. 
The risk of LLMs not only generating but potentially magnifying existing social biases is a matter of concern as well~\cite{cao2023comprehensive}.

\smallskip
\noindent
\textbf{Theme 7. Response Format and Attitude (\format{})}
Dissatisfaction with the format of responses, including tone, length, structure, and overall attitude, was captured within this theme. This dissatisfaction can arise when users have expectations regarding the manner in which responses were delivered and the tone used by the LLM.
ChatGPT's choice of words and formal, dry tone ~\cite{guo2023close}, as well as extensive and detailed responses ~\cite{yang2023exploring} are quite different from human-generated text, which was colloquial and shorter.

These seven themes collectively offer a structured framework for understanding the multifaceted nature of user dissatisfaction with ChatGPT responses. Our survey utilized these themes as a basis for systematically investigating and quantifying user dissatisfaction.

\section{Data Collection} \label{sec:data}
Based on the categorization of user-side dissatisfaction from LLM responses, we collected the actual user's ChatGPT conversation log data with the dissatisfaction through a data collection system we designed and implemented. 
Our system targeted individuals who have utilized ChatGPT for practical purposes such as increasing productivity or efficiency in work, study, or hobbies.
This process aims to address the following three research questions: 
\smallskip
\begin{enumerate}
    \item[\textbf{RQ1.}] What and How much dissatisfaction do users experience from LLM-generated responses?
    \item[\textbf{RQ2.}] How do users address these dissatisfactions in their subsequent prompts during the conversation with LLM?
    \item[\textbf{RQ3.}] How do user dissatisfaction and tactics vary depending on users' knowledge level regarding LLMs?
\end{enumerate}
\begin{figure*}[!ht]
    \centering    \includegraphics[width=\textwidth]{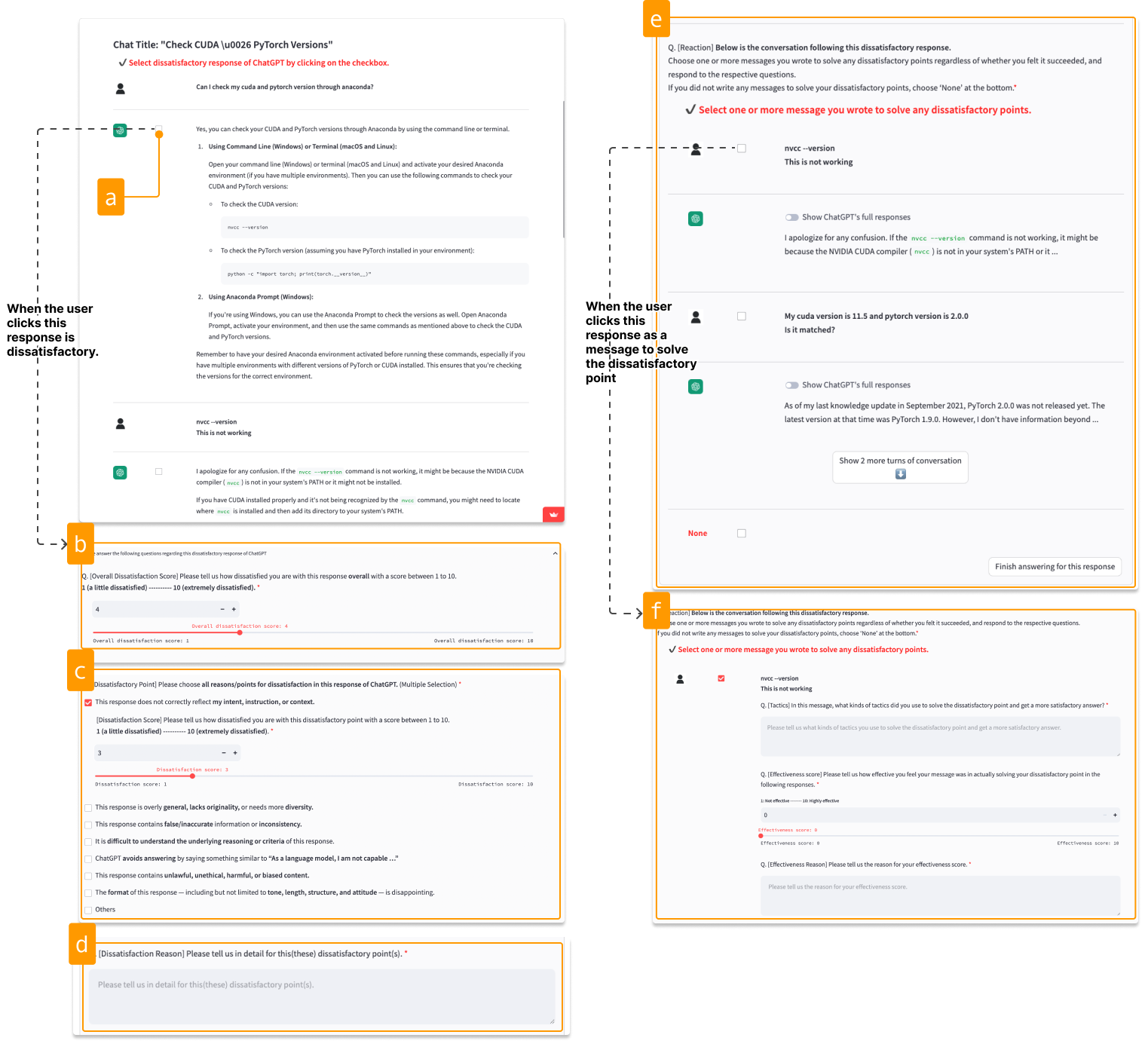}
    \caption{Screenshot of the data collection system.
    }
    \Description{This figure illustrates the stage 4 of the data collection system. The details can be found in Stage 4. Answering Questions about dissatisfactory responses of Section 4.1 Data collection System Design.}
    \label{fig:system}
\end{figure*}
\subsection{Data Collection System Design}
To collect users' ChatGPT conversation log data in the wild, we designed and implemented a data collection system that includes the following four stages.   

\noindent
\textbf{Stage 1. Answering a General Questionnaire}
In the first stage, we collected demographic information of participants such as gender, age, occupation, and overall experiences with ChatGPT (e.g., the frequency and period of using ChatGPT in their workflow).
We also asked about the participants' knowledge level regarding Large Language Models (LLM) (``Regarding the mechanisms of Large-language models such as ChatGPT, how much do you agree with the following statement?''). All questions in this stage were measured through a 7-point Likert scale. 

\noindent
\textbf{Stage 2. Looking Through ChatGPT Chat History}
In stage 2, participants were instructed to review their ChatGPT conversation history that had happened within 30 days. While reviewing, we asked the participants to find a conversation in which they experienced dissatisfaction with ChatGPT responses. To facilitate participants to think of various cases of dissatisfaction, we provided the descriptions of dissatisfaction categories derived from our systematic literature review as examples.

\noindent
\textbf{Stage 3. Submitting Dissatisfactory Conversations}
Based on their reflections regarding dissatisfaction in stage 2, we requested the participants to share a ChatGPT conversation link ~\footnote{https://help.openai.com/en/articles/7943611-create-a-shared-link} within the past 30 days in which they experienced at least one dissatisfactory response. The participants can input the link into our system. To collect the conversation data with the details of the context, we also asked them to provide information about the purpose of the conversation, the reasons for using ChatGPT in that context, and the version of ChatGPT they used in this conversation, like GPT-3.5. Lastly, we asked the participants how much they remembered the conversation.

\noindent
\textbf{Stage 4. Answering Questions About Dissatisfactory Responses}
The participant's shared link was processed by transforming ChatGPT responses and user prompts to be presented as selectable components in the system (Fig \ref{fig:system}-a). The system also allowed participants to provide specific experiences of dissatisfactory responses by selecting each response (Fig \ref{fig:system}-b\textasciitilde f). 
For each selected response, participants were asked to 
(1) rate the overall level of dissatisfaction on a scale of 1 to 10 (1: a little dissatisfied, 10: extremely dissatisfied) (Fig \ref{fig:system}-b),
(2) choose one or more dissatisfaction categories from the given seven categories, or optionally describe a custom dissatisfaction point for dissatisfaction (Fig \ref{fig:system}-c),
(3) rate the level of dissatisfaction for each selected category on a scale of 1 to 10 (1: a little dissatisfied, 10: extremely dissatisfied) (Fig \ref{fig:system}-c), 
(4) provide a detailed free-form explanation for their dissatisfaction (Fig \ref{fig:system}-d), 
(5) select a prompt among the subsequent conversations in which they tried to resolve the dissatisfaction (Fig \ref{fig:system}-e), 
(6) describe their tactic to address the dissatisfaction in the prompt (Fig \ref{fig:system}-f),
(7) rate the effectiveness of their tactic on a scale of 1 to 10 (1: not effective, 10: highly effective) (Fig \ref{fig:system}-f),
(8) provide a written explanation of the reasons for their effectiveness rating (Fig \ref{fig:system}-f). In cases where there was no subsequent prompt or the conversation ended after dissatisfaction, participants were asked to provide written reasons instead of responding to (5)-(8).

\subsection{Collected Data}
\subsubsection{Participants and Collected Data}
We distributed the data collection system to people over the age of 18 globally through the Prolific platform~\footnote{https://www.prolific.co/}. 
Participants who provided at least two ChatGPT conversation links and evaluated at least one dissatisfactory response for each link received a compensation of £6.
For each additional dissatisfactory response submitted from a single conversation link, participants received an additional £0.75 per response. 
For each additional conversation link provided beyond the initial two, participants received an additional £1.5 per link.
We limited the number of maximum conversation links that can be submitted to five for each participant to prevent one participant from providing lots of conversation links.
In total, we collected 307 ChatGPT conversation links, 511 dissatisfactory ChatGPT responses, and 615 user responses regarding those dissatisfactions from 107 individuals.
Each user submitted an average of 2.87 links (std=1.21), 4.78 dissatisfactory ChatGPT responses (std=5.61), and 5.75  responses regarding those dissatisfactions (std=6.62).
This study was approached by our institution's IRB, and we received consent from participants for the release of datasets.

\subsubsection{Data Filtering and Pre-processing} \label{sec:data-processing}
To ensure the quality and reliability of the data collected from our system, two authors reviewed all the data together according to the following criteria and conducted filtering or pre-processing where necessary. 

\smallskip
\noindent
\textbf{Filtering Process}
The data was filtered out at three levels: (1) user, (2) conversation, and (3) dissatisfactory responses.

\noindent
\textbf{\textit{1. User-Level Filtering}}
We identified that one participant provided altogether contradictory responses, which contradicted the dissatisfactory response and the effectiveness of the prompt in resolving the dissatisfaction.
Consequently, all data from this user were excluded.

\noindent
\textit{\textbf{2. Conversation-Level Filtering}}
The conversation-level filtering was conducted based on the following four criteria, and a total of 20 conversations were filtered out. The detailed reason for each criteria is in the Appendix (Sec~\ref{appendix:criteria}). 
\begin{enumerate}
    \item Conversation older than 30 days.
    \item Conversation with a memory level of 3 or lower.
    \item Conversation for fun or testing purposes.
    \item Conversation from versions other than GPT-3.5.
\end{enumerate}

\noindent
\textbf{\textit{3. Response-Level Filtering}}
Response-level filtering was conducted based on the following four criteria, leading to the exclusion of a total of 16 dissatisfactory ChatGPT responses. 
\begin{enumerate}
    \item Dissatisfaction due to ChatGPT's error messages
    \item Unconvincing dissatisfaction
    \item Mismatch between score and reason
    \item No Correlation between selected dissatisfactions and subsequent prompts for resolving that dissatisfaction
\end{enumerate}

Detailed reasons and examples of each filtering case can be found in the Appendix and supplementary material. Please note that when filtering at the response level, all associated subsequent prompts and tactic data related to that response were also filtered. When filtering at the conversation level, all data related to the ChatGPT dissatisfactory responses and user prompts within that conversation were also filtered out. When filtering at the user level, all data provided by that user were excluded.

\smallskip
\noindent
\textbf{Pre-processing Process}
The data pre-processing process primarily involved the reassignment of dissatisfaction categories. This step was undertaken to deal with cases where participants incorrectly selected dissatisfaction categories or opted for the `other' option when evaluating the dissatisfaction category. Two authors examined all the data and carried out reassignment according to the following two criteria, proceeding only when a consensus was reached. Detailed examples of each case where reassignment occurred can be found in the supplemental.

Criterion 1: Reassigning `other' to a specific category. For the `other' option, when we found that there was a more suitable match with another category that was not selected based on the dissatisfaction reason, the `other' score was reallocated to the corresponding category. 
As a result of this criterion, four entries were reassigned to the \intent{} category, two to \depth{}, three to \acc{}, and five to \format{}.

Criterion 2: Reassigning an incorrectly selected category to another. If a participant had only checked one dissatisfaction category, and upon reviewing the dissatisfaction reason and conversation, it was evident that the selected category was not appropriate but another category was a better fit, the score was reassigned to the more suitable category.
Using this criterion, three entries were reallocated from \intent{} to \format{}, three from \intent{} to \acc{}, two from \acc{} to \intent{}, one from \acc{} to \depth{}, and two from \depth{} to \format{}.

\subsubsection{Dataset}
After filtering and pre-processing, we built a dataset on end-users' dissatisfaction with ChatGPT and their responses.
The dataset is hierarchically organized, comprising the following components:

\begin{enumerate}
    \item User (N=94)
    \item ChatGPT conversation links and logs (N=249)
    \item User's recollected experience data on dissatisfactory ChatGPT responses (N=377)
    \item User's strategies to respond to the dissatisfactory response (N=459)
\end{enumerate}
\noindent
Here, the user's strategies were qualitatively analyzed, resulting in the creation of 13 tactic codes categorized into four themes. 
More detail of this is in Sec ~\ref{sec:result-2}. Each data is also labeled as corresponding tactic codes by the authors.
With this dataset, we conducted a quantitative and qualitative analysis to answer our research questions. 
We provide this dataset to facilitate future research about user experiences on chat-based LLMs.
In releasing the dataset, we took careful consideration by masking all sensitive information related to their privacy and personal information.
A more detailed description about the dataset can be accessed through our project website~\footnote{https://chatgpt-analysis.kixlab.org}.
\section{Data Analysis and Results}
In this section, we present the analysis method and results that answer our research questions based on the constructed dataset.
Firstly, we present the analysis of the types of dissatisfaction users face in LLM responses (RQ1).
Next, we present how users respond to dissatisfaction through qualitative analysis (Table ~\ref{tab:tactic-result}) and analyze the effectiveness of the tactics users use (RQ2).
Finally, we present how users' knowledge level regarding LLM influences their experiences of dissatisfaction and their behaviors when they face dissatisfaction (RQ3). 

\begin{table*}[!ht]
{\small
\centering
\resizebox{0.75\textwidth}{!}{%
\def\arraystretch{1.2}%
\begin{tabular}{c|cc|c}
\toprule
\multirow{2}{*}{\makecell[c]{Dissatisfaction \\ Category}} &
  \multicolumn{2}{c|}{Response-level analysis} & 
  \multicolumn{1}{c}{User-level analysis} \\ \cline{2-4} 
 &
  Count: N (\%) &
  \makecell[c]{Dissatisfaction Score: mean (std)*} &
  \multicolumn{1}{c}{Frequency: mean (std)*} \\ 
\midrule
\intent{}    & \textbf{168 (32.18\%)} & 5.56 (2.94)            & \textbf{0.47 (0.03)} \\
\depth{}     & 107 (20.50\%)          & \textbf{5.09 (2.69) *} & 0.33 (0.35) \\
\acc{}       & 83 (15.90\%)           & \textbf{6.52 (2.76) *} & 0.20 (0.03) \\
\trans{}  & 27 (5.17\%)            & 4.81 (3.13)            & 0.08 (0.02)\\
\refuse{} & 27 (5.17\%)            & 6.37 (2.68)            & 0.09 (0.02) \\
\ethic{} & 4 (0.77\%)             & 6.25 (3.20)            & 0.01 (0.01) \\
\format{} & 106 (20.31\%)          & 6.14 (3.04)            & 0.27 (0.03) \\
\bottomrule
\end{tabular}%
}
\caption{Analysis results on the count, dissatisfaction score, and user-level frequency for the dissatisfaction category (* p-value < 0.01)}
\Description{This table depicts the analysis results on the count, dissatisfaction scores, and user-level frequency analysis for the dissatisfaction category.}
\label{tab:dis}
}
\end{table*}
\subsection{RQ1. Analysis of how users experience dissatisfaction}
\subsubsection{Dissatisfaction Category Analysis}
We analyzed the count, distribution, and dissatisfaction score of the seven categories of dissatisfaction organized through a systematic literature review in Section~\ref{sec:slr}, and the results are described in Table ~\ref{tab:dis}. 
In terms of the count of each category, \intent{} accounted for the largest proportion (32.18\%), while \trans{}, \refuse{}, and \ethic{} constituted significantly smaller proportions compared to the other categories. 
To investigate the severity degree of user dissatisfaction in each category, we conducted \kw{} test and confirmed significant differences between categories ($\chi^2$ = 17.6, p-value < 0.01, df = 6). In particular, we found that \acc{}'s dissatisfaction score was the highest, and its score was statistically significantly higher than \depth{} through Dwass-Steel-Critchlow-Fligner(DSCF) pairwise comparison (p-value=0.008).
This means that users are statistically significantly more dissatisfied with dissatisfaction due to \acc{} than \depth{}.

Considering that each user provided multiple dissatisfactory responses, we also conducted a user-level analysis, accounting for potential correlations among the data submitted by the same user. 
To achieve this, we normalized each dissatisfaction category data by dividing them by the number of dissatisfactory responses each user submitted. 
This method allowed us to express each data point as the frequency of how often each user experienced dissatisfaction in a certain category.
The analysis results are presented in Table \ref{tab:dis} in the ``User-level'' analysis column.
The mean frequency value of \intent{} was 0.47, indicating that if a user has experienced 100 dissatisfactory ChatGPT responses, on average, 47 of them fall into the \intent{} category.
Furthermore, the \kw{} test result shows statistically significant differences in user-level frequency values between each category ($\chi^2$ = 9.93, p-value < 0.01, df = 6).
In the user-level analysis, we can see a similar tendency to the response-level analysis, users experience \intent{} the most frequently.
Following this, the second most frequently encountered dissatisfaction is \depth{}. However, the standard deviation of \depth{} is 0.35, which is much higher than other categories, indicating that the frequency of experiencing \depth{} varies significantly from user to user. 

\begin{figure*}[!ht]
    \centering
    \includegraphics[width=\textwidth]{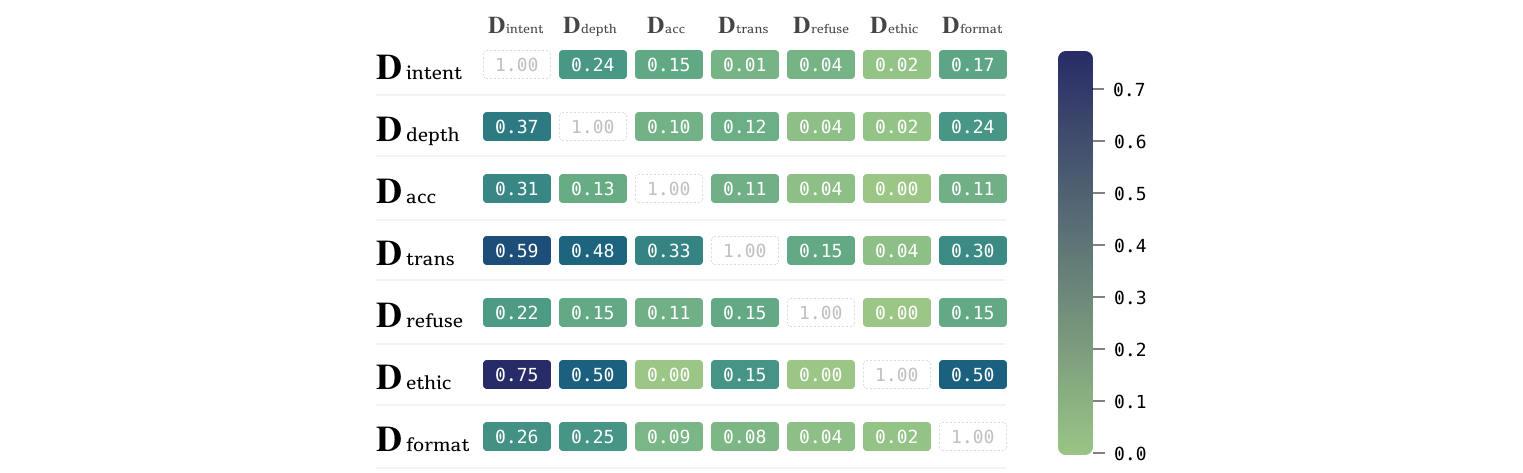}
    \caption{Normalized Co-occurrence matrix of dissatisfaction category. The value at (\textit{i}, \textit{j}) in this matrix represents the frequency of when the \textit{i}th row was selected as a dissatisfaction point, the \textit{j}th column was also selected as a dissatisfaction.
    }
    \Description{This figure is a co-occurrence matrix of dissatisfaction categories. It is a seven-by-seven matrix. For each row of dissatisfaction category i, the seven values are the frequency that each dissatisfaction category j was selected when i was selected by the user.}
    \label{fig:co_occr_nor}
\end{figure*}
        
\subsubsection{Co-occurrence Analysis}
In a single dissatisfactory response, multiple dissatisfaction categories can co-occur. 
For example, a user may simultaneously experience dissatisfaction with the lack of originality (\depth{}) and the length (\format{}) of ChatGPT's response at the same time. 
Therefore, we analyzed co-occurrence patterns to investigate the correlations between each category of dissatisfaction.
Results are presented in Fig ~\ref{fig:co_occr_nor} and the value at (\textit{i}, \textit{j}) in this matrix represents the frequency of when the \textit{i}-th row was selected as a source of dissatisfaction, the \textit{j}-th column was also selected together. The result shows that \intent{} frequently appears concurrently with all other categories. Also, while \trans{} and \ethic{} have relatively low counts, they co-occur with \intent{} more than half the times in each occurrence.

\subsection{RQ2. Analysis of how users respond to dissatisfaction}\label{sec:result-2}
\begin{table*}[!ht]
\tiny{
\begin{center}
\resizebox{\linewidth}{!}{%
\def\arraystretch{1.25}%
\begin{tabular}{p{0.2\columnwidth}p{0.8\columnwidth}}
\toprule
    \textbf{Category (4)} & \textbf{Tactic Code (13)} \\
\midrule
\multirow{3}{*}{\makecell[l]{\\Prompt Reusing \\ and Shortcut\\(\Trepeat{})}} & T1: Re-using an identical prompt or slightly paraphrasing it \\
& T2: Using the specific word (e.g., more, another) that implies requesting different or more outputs for the same task as the previous prompt \\ 
& T3: Re-using an identical prompt but adding emphasis through formatting (e.g., using all capital letters, using double quotation marks) \\
\hline

\multirow{4}{*}{\makecell[l]{Intent Concretization\\(\Tspecify{})}} & T4: Specifying user intent by providing detailed or direct instructions \\ 
& T5: Specifying user intent by providing additional context or explanation \\ 
& T6: Adding format-specific conditions (e.g., make it shorter, provide in list format) \\
& T7: Adding tone-specific conditions (e.g., make it casual) \\
\hline

\multirow{3}{*}{\makecell[l]{Error Identification\\and Correction\\(\Terror{})}} & T8: Pointing out errors or mistakes \\ 
& T9: Providing the correct answer or hints \\ 
& T10: Asking clarification questions \\
\hline

\multirow{3}{*}{\makecell[l]{Task Adaptation\\(\Tadapt{})}} & T11: Adapting by shifting to another topic or task that is different from the original intent. \\ 
& T12: Breaking down the original task into smaller subtasks \\ 
& T13: Asking follow-up questions deviating from the original task \\
\hline
\makecell[l]{\Tno{}} & No further prompting to address the dissatisfaction and even terminating the conversation due to dissatisfaction \\
\bottomrule
\end{tabular}%
}
\end{center}
}
\caption{User tactic category}
\Description{This table summarized the user tactic categories found with qualitative analysis of user data. There are two columns: Category and Tactic Code. There are four tactic categories and additional category which is No Tactic. Each tactic category has detailed tactic codes. The details can be found in Section 5.2.}
\label{tab:tactic-result}
\end{table*}

\subsubsection{Categorizing Tactics for Resolving Dissatisfaction}
Through qualitative analysis, we categorized users' tactics to understand and analyze how users address their dissatisfaction from ChatGPT's response through subsequent prompts. 
Two authors independently conducted open coding by reviewing ChatGPT conversation log data, user-side dissatisfactions on ChatGPT responses, employed tactics in subsequent prompts, and user-reported effectiveness and the reasons for these tactics. After completing the open coding, the two authors engaged in an iterative process of code consolidation. To precisely capture and categorize the subtleties of user tactics, both authors iterated all data together, making a code set through discussion. We proceeded with these processes until the authors met a common ground. After two times of iterations, we identified the user's tactic with 13 codes as presented in Table~\ref{tab:tactic-result}. To establish relationships between these codes and identify overarching themes, axial coding ~\cite{strauss1998basics} was performed. Through this coding process, we identified four main themes of the user's tactics, as presented in Table ~\ref{tab:tactic-result}.

\noindent
\textbf{Tactic Category 1. Prompt Reusing and Shortcut} 
This category of tactic represents users either reusing prompts or employing a single word to request similar or diverse responses, often requiring minimal effort in crafting the prompt.
This category comprises three tactics. First, users just reuse the exact same prompt as the previous one or paraphrase it slightly (T1). Second, users use a single word like `more' or `another' as a shortcut to get either similar responses from the previous turn or a wider range of responses from ChatGPT (T2).
Last, users retry by adding emphasis through formatting, such as using all capital letters or using double quotation marks (T3). 

\noindent
\textbf{Tactic Category 2. Intent Concretization} 
This category encompasses four tactics of users trying to concretize their intent and context to get a more appropriate response.
Users further specify their needs by providing more detailed or direct instructions (T4), giving additional context or explanation (T5).
For example, if users ask ChatGPT to recommend a dinner menu and they doesn't like ChatGPT's answer, they can further specify their needs by saying, ``Recommend a \textbf{healthy} dinner menu using \textbf{tomatoes}'' (T4), or explain their context by saying, ``I'm going to invite a guest to my house for my dinner'' (T5).
And users concretize their intent by adding specific conditions related to the format such as ``make it shorter'' (T6), and adding specific conditions related to the tone, such as ``make it casual'' (T7).

\noindent
\textbf{Tactic Category 3. Error Identification and Correction} 
This category mainly contains tactics when there are some errors in the ChatGPT's response, and the users point out or correct them.
Users simply say ``It was wrong.'' or point out the part that is wrong (T8), give the correct answer or hints of the correct answer (T9), and ask a clarification question to confirm the error or doubtful aspects such as by asking ``Can you confirm that ... ?'' or ``Are you sure ...?''(T10). 

\noindent
\textbf{Tactic Category 4. Task Adaptation} 
This category represents the user adjusting to another task instead of the original task where the user felt dissatisfied. 
Users adapt their task by altering their initial task to a different one (T11). For instance, if users initially ask for the latest information and ChatGPT says it can only answer up to 2021 information, then they can slightly adjust their original task and ask for 2021 information rather than the latest information.
Users also adjust their original task by dividing it into smaller and more manageable subtasks (T12). For example, when users ask ChatGPT for a complex math problem, they can ask them in intermediate steps.
Finally, Users ask follow-up questions deviating from the original task, such as asking follow-up questions about parts that lack details or are unfamiliar to them in ChatGPT's responses. (T13). 

\subsubsection{Tactic Category Analysis}
After creating the tactic categories, we categorized users' prompts into four tactic categories or \Tno{}. 
\Tno{} indicates no further prompting to address the dissatisfaction and even terminating the conversation due to dissatisfaction. 
Here, note that a single user prompt can encompass multiple tactic categories if the prompt contains multiple requests.
We conducted response-level analysis for the count, distribution, and effectiveness of each tactic as well as user-level analysis for frequency (Table ~\ref{tab:tactic_analysis}).
Notably, we observed that \Tspecify{} stands out as the dominant category, and it accounts for over half of the distribution (58.6\%) among the four tactic categories without \Tno{}. 
In addition, we analyzed the effectiveness of each tactic based on users' rating of the effectiveness score between 1 and 10. 
We conducted a \kw{} test and confirmed that there are statistically significant differences between the effectiveness scores of each tactic ($\chi^2$ = 23.1, p-value < 0.01, df = 4). 
Specifically, we found that \Tspecify{}, a tactic for users to further specify their own intents, is most effective with a mean score 0f 6.04, highest of all categories.
\begin{table*}[!h]
\centering
\resizebox{0.80\textwidth}{!}{%
\def\arraystretch{1.15}%
\begin{tabular}{cc|cccccccc|cccc}
\toprule
\multirow{3}{*}{\begin{tabular}[c]{@{}c@{}}Tactic \\ Category\end{tabular}} &
  \multirow{3}{*}{\begin{tabular}[c]{@{}c@{}}Tactic \\ Code\end{tabular}} &
  \multicolumn{8}{c|}{Response-level analysis} &
  \multicolumn{4}{c}{User-level analysis} \\ \cline{3-14} 
 &
   &
  \multicolumn{4}{c|}{Count:  N (\%)} &
  \multicolumn{4}{c|}{Effectiveness Score: mean (std)} &
  \multicolumn{4}{c}{Frequency: mean (std)} \\ \cline{3-14} 
 &
   &
  \multicolumn{2}{c}{Category} &
  \multicolumn{2}{c|}{Code} &
  \multicolumn{2}{c}{Category*} &
  \multicolumn{2}{c|}{Code} &
  \multicolumn{2}{c}{Category} &
  \multicolumn{2}{c}{Code} \\ 
\midrule
\multirow{3}{*}{\Trepeat{}} & {T1} &
  \multicolumn{2}{c}{\multirow{3}{*}{45 (9.4\%)}} &
  \multicolumn{2}{c|}{29 (5.8\%)} &
  \multicolumn{2}{c}{\multirow{3}{*}{4.04 (3.16)}} &
  \multicolumn{2}{c|}{4.45 (3.15)} &
  \multicolumn{2}{c}{\multirow{3}{*}{0.09 (0.20)}} &
  \multicolumn{2}{c}{0.07 (0.18)} \\
 &  {T2} &
  \multicolumn{2}{c}{} &
  \multicolumn{2}{c|}{18 (3.6\%)} &
  \multicolumn{2}{c}{} &
  \multicolumn{2}{c|}{3.06 (3.06)} &
  \multicolumn{2}{c}{} &
  \multicolumn{2}{c}{0.02 (0.09)} \\
 &  {T3} &
  \multicolumn{2}{c}{} &
  \multicolumn{2}{c|}{2 (0.4\%)} &
  \multicolumn{2}{c}{} &
  \multicolumn{2}{c|}{1.00 (0.00)} &
  \multicolumn{2}{c}{} &
  \multicolumn{2}{c}{0.00 (0.04)} \\ \hline
\multirow{4}{*}{\Tspecify{}} &  {T4} &
  \multicolumn{2}{c}{\multirow{4}{*}{\textbf{183 (38.4\%)}}} &
  \multicolumn{2}{c|}{122 (24.4\%)} &
  \multicolumn{2}{c}{\multirow{4}{*}{\textbf{6.04 (3.44)}}} &
  \multicolumn{2}{c|}{6.25 (3.53)} &
  \multicolumn{2}{c}{\multirow{4}{*}{\textbf{0.33 (0.34)}}} &
  \multicolumn{2}{c}{0.22 (0.28)} \\
 &  {T5} &
  \multicolumn{2}{c}{} &
  \multicolumn{2}{c|}{26 (5.2\%)} &
  \multicolumn{2}{c}{} &
  \multicolumn{2}{c|}{5.35 (3.33)} &
  \multicolumn{2}{c}{} &
  \multicolumn{2}{c}{0.06 (0.15)} \\
 &  {T6} &
  \multicolumn{2}{c}{} &
  \multicolumn{2}{c|}{40 (8.0\%)} &
  \multicolumn{2}{c}{} &
  \multicolumn{2}{c|}{6.45 (3.16)} &
  \multicolumn{2}{c}{} &
  \multicolumn{2}{c}{0.08 (0.17)} \\
 &  {T7} &
  \multicolumn{2}{c}{} &
  \multicolumn{2}{c|}{11 (2.2\%)} &
  \multicolumn{2}{c}{} &
  \multicolumn{2}{c|}{4.73 (3.04)} &
  \multicolumn{2}{c}{} &
  \multicolumn{2}{c}{0.02 (0.10)} \\ \hline
\multirow{3}{*}{\Terror{}} &
  {T8} &
  \multicolumn{2}{c}{\multirow{3}{*}{73 (15.3\%)}} &
  \multicolumn{2}{c|}{53 (10.6\%)} &
  \multicolumn{2}{c}{\multirow{3}{*}{4.19 (2.95)}} &
  \multicolumn{2}{c|}{4.26 (2.99)} &
  \multicolumn{2}{c}{\multirow{3}{*}{0.10 (0.22)}} &
  \multicolumn{2}{c}{0.06 (0.16)} \\
 &  {T9} &
  \multicolumn{2}{c}{} &
  \multicolumn{2}{c|}{13 (2.6\%)} &
  \multicolumn{2}{c}{} &
  \multicolumn{2}{c|}{4.62 (2.66)} &
  \multicolumn{2}{c}{} &
  \multicolumn{2}{c}{0.02 (0.09)} \\
 &  {T10} &
  \multicolumn{2}{c}{} &
  \multicolumn{2}{c|}{10 (2.0\%)} &
  \multicolumn{2}{c}{} &
  \multicolumn{2}{c|}{3.80 (3.16)} &
  \multicolumn{2}{c}{} &
  \multicolumn{2}{c}{0.03 (0.10)} \\ \hline
\multirow{3}{*}{\Tadapt{}} &
  {T11} &
  \multicolumn{2}{c}{\multirow{3}{*}{12 (2.5\%)}} &
  \multicolumn{2}{c|}{7 (1.4\%)} &
  \multicolumn{2}{c}{\multirow{3}{*}{5.17 (3.04)}} &
  \multicolumn{2}{c|}{4.57 (3.21)} &
  \multicolumn{2}{c}{\multirow{3}{*}{0.04 (0.11)}} &
  \multicolumn{2}{c}{0.03 (0.10)} \\
 &
  {T12} &
  \multicolumn{2}{c}{} &
  \multicolumn{2}{c|}{2 (0.4\%)} &
  \multicolumn{2}{c}{} &
  \multicolumn{2}{c|}{8.00 (0.00)} &
  \multicolumn{2}{c}{} &
  \multicolumn{2}{c}{0.00 (0.03)} \\
 &
  {T13} &
  \multicolumn{2}{c}{} &
  \multicolumn{2}{c|}{3 (0.6\%)} &
  \multicolumn{2}{c}{} &
  \multicolumn{2}{c|}{4.67 (3.22)} &
  \multicolumn{2}{c}{} &
  \multicolumn{2}{c}{0.00 (0.04)} \\ \hline
\multicolumn{2}{c|}{\Tno{}} &
  \multicolumn{2}{c}{164 (34.4\%)} &
  \multicolumn{2}{c|}{164 (32.8\%)} &
  \multicolumn{2}{c}{-} &
  \multicolumn{2}{c|}{-} &
  \multicolumn{2}{c}{0.47 (0.38)} &
  \multicolumn{2}{c}{0.47 (0.36)} \\
\bottomrule
\end{tabular}%
}
\caption{Analysis results on the count, effectiveness score, and user-level frequency for the tactic category
(* p-value < 0.01)}
\Description{This table shows the analysis results on the count and effectiveness scores on Tactic categories and codes. For the four Tactic categories and No Tactic, as well as 19 tactic codes, the count and percentage, effectiveness score mean and standard deviation is shown. Also, we calculate the user-level frequency for the tactics and show their mean and standard deviation.}
\label{tab:tactic_analysis}
\end{table*}

\begin{figure*}[t]
    \centering
    \subfigure[]{
        \includegraphics[width=0.42\textwidth]{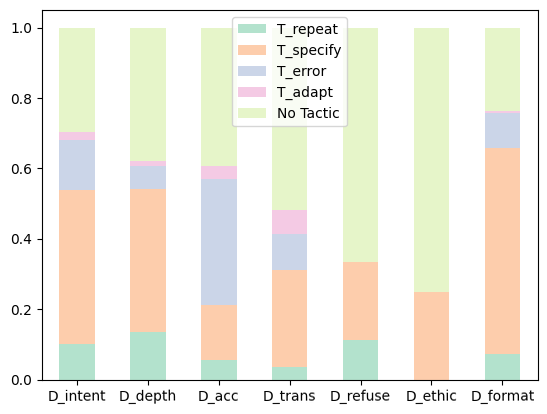}
        \label{fig:dis_tactic_bar}
    }
    \subfigure[]{
        \includegraphics[width=0.5\textwidth]{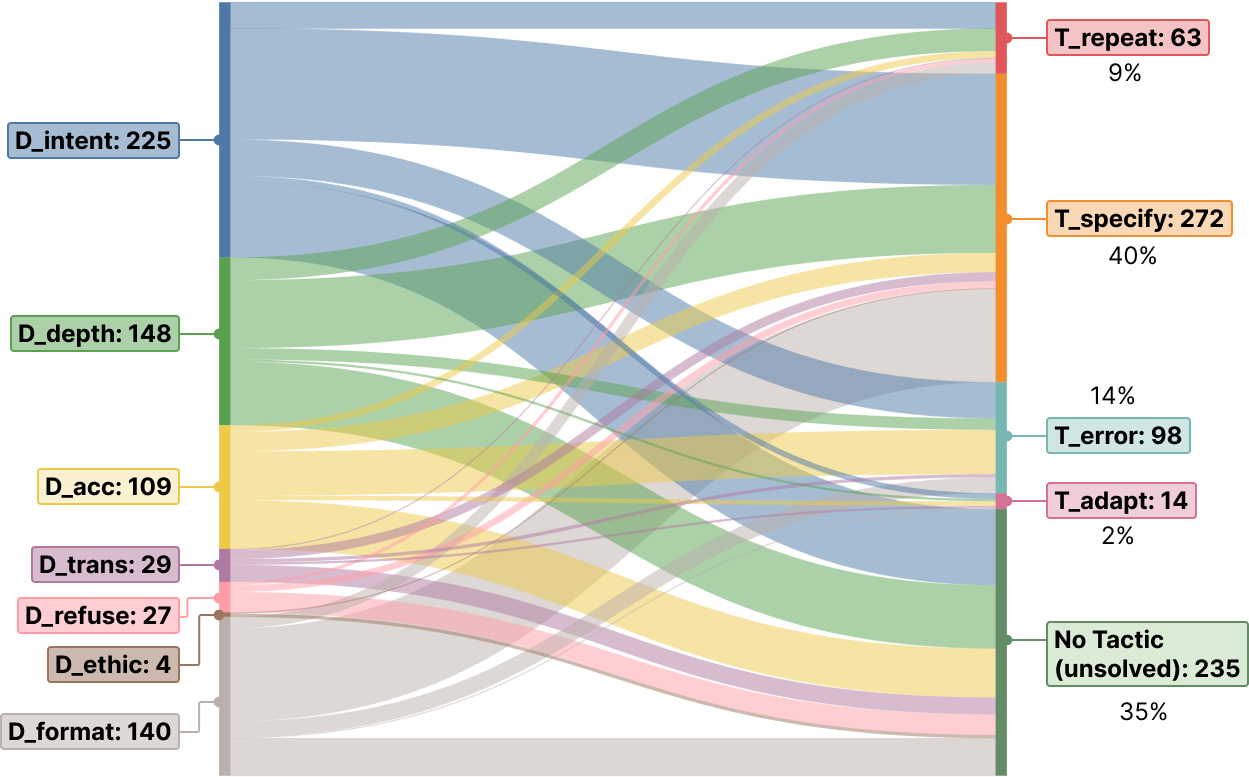}
        \label{fig:sankey_12}
    }
\caption{ 
(a) Distribution of tactic categories by dissatisfaction category.
(b) Sankey diagram to visualize how users respond among four tactic categories or \Tno{} after experiencing each of the dissatisfaction categories. Note that the count in the Sankey diagram can be greater than the count of response-level analysis in Table ~\ref{tab:dis} and ~\ref{tab:tactic_analysis}. This is because one response can include multiple dissatisfaction categories and multiple tactic categories, and they were counted multiple times to draw a Sankey diagram.
}
\Description{There is one stacked bar chart (a) and one sankey diagram (b). (a) is a stacked bar chart showing the distribution of four tactic categories and No Tactic for each of seven dissatisfaction categories. T_error has the highest proportion in D_acc, while T_specify is most used in D_format, followed by D_intent and D_depth. T_adapt is the least used tactic in all dissatisfaction categories. (b) is a sankey diagram depicting the flow from dissatisfaction categories to tactic categories or No Tactic. It addresses the same data as (a) but in a different style}
\label{fig:dis-tactic}
\end{figure*}

\subsubsection{Dissatisfaction Category and Corresponding Tactics: Whether the dissatisfaction was solved}
We investigated how users applied different tactics to address each dissatisfaction category and whether these tactics resolved the dissatisfaction.
Firstly, we analyzed the distribution of tactics used for each dissatisfaction category (Fig. \ref{fig:dis_tactic_bar}), and drew a Sankey diagram to visualize the overall flow of tactics used by each dissatisfaction category (Fig. \ref{fig:sankey_12}).
We observed that \Tspecify{} is the dominant tactic across various dissatisfaction categories.
However, when users encounter dissatisfaction related to the accuracy of information (\acc{}), they tend to employ \Terror{} rather than \Tspecify{}. 
Lastly, in cases of \trans{}, \refuse{}, and \ethic{}, users often resort to \Tno{}, ending up the conversation.
\begin{figure*}[h]
    \centering
    \subfigure[]{
        \includegraphics[width=0.43\textwidth]{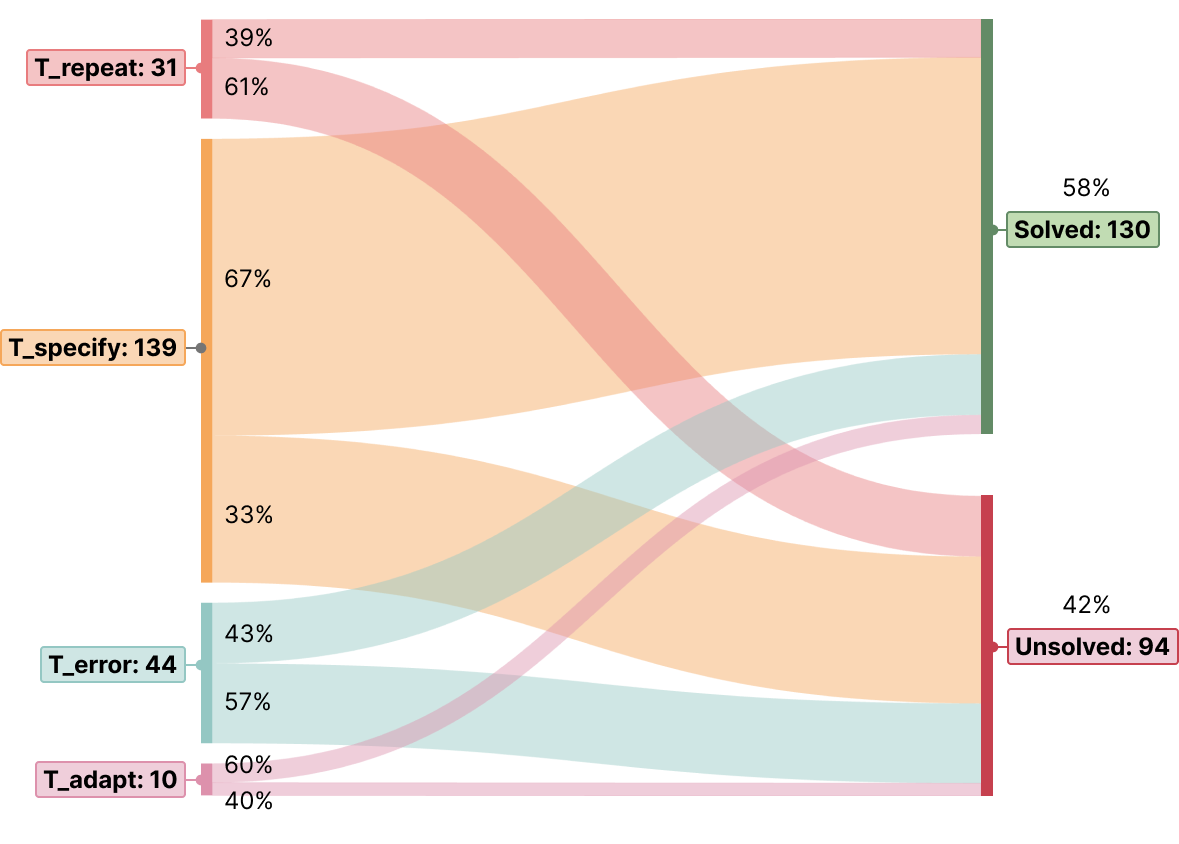}
        \label{fig:sankey23}
    }
    \subfigure[]{
        \includegraphics[width=0.54\textwidth]{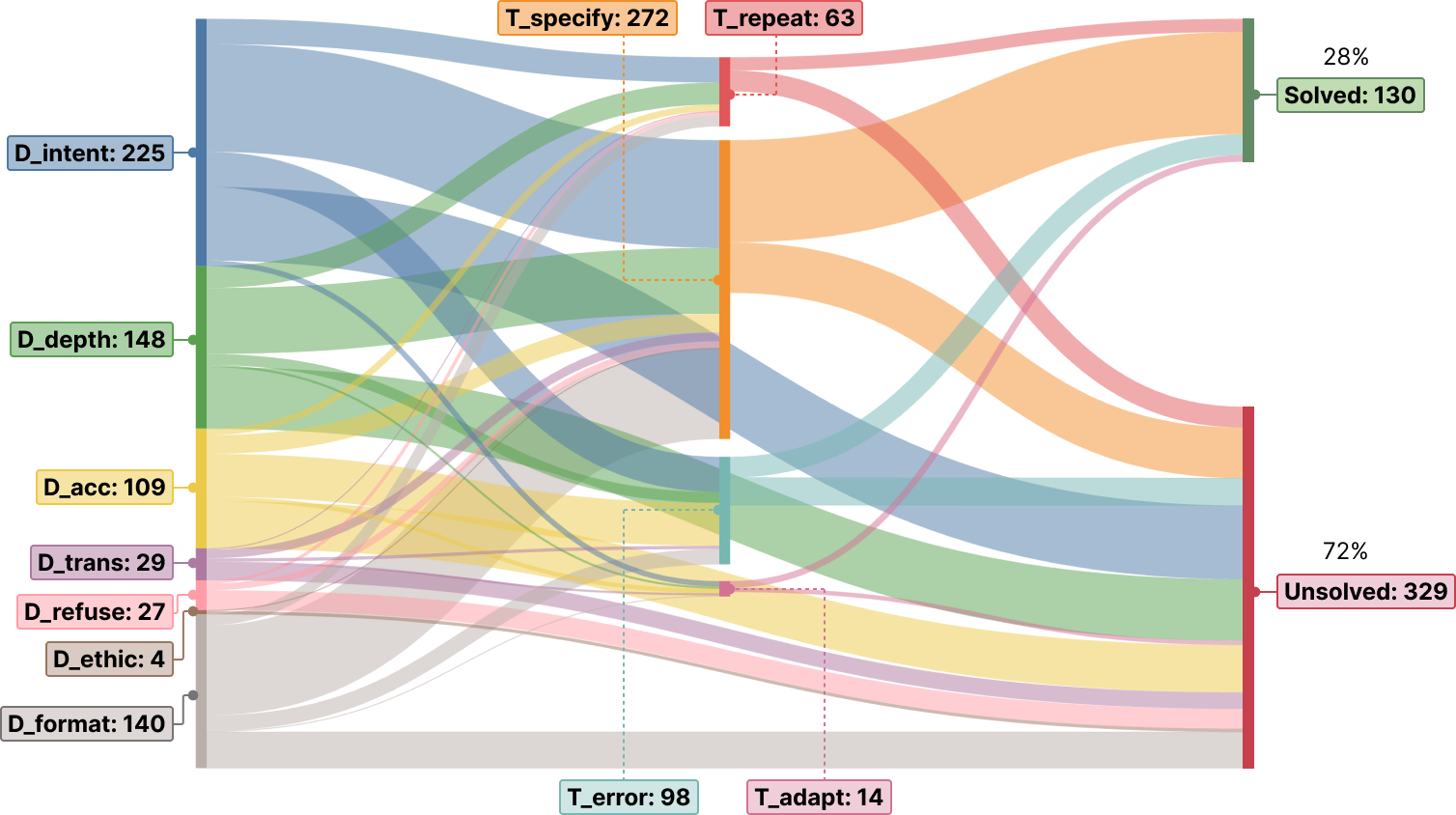}
        \label{fig:sankey123}
    }
\caption{
(a) A Sankey diagram that visualizes whether users resolved their dissatisfaction using each of the tactic categories.
(b) The overall visualization of how users respond among the four tactic categories after experiencing each of the dissatisfaction categories and finally whether that dissatisfaction was solved or not. 
}
\Description{There are two figures and all of them are Sankey diagrams to visualize how users respond to each of the dissatisfaction categories using which tactic, and finally whether it was solved or not. (a): It shows whether users resolved their dissatisfaction using different tactic categories. (b): It is the overall visualization combining these aspects to provide a holistic view of user interactions with ChatGPT.}
\end{figure*}
The proportion and visualization of whether or not dissatisfaction has been resolved by each tactic can be seen in Fig. ~\ref{fig:sankey23}.
Fig. ~\ref{fig:sankey23} illustrates that the users managed to resolve their dissatisfaction by 58 \% by utilizing tactics. 
Notably, \Tspecify{} was an effective way of resolving dissatisfaction in many cases (67\%), while with other tactics, there were more cases where dissatisfaction remained unsolved.
Fig. ~\ref{fig:sankey123} shows which tactics users use for each dissatisfaction category and how this eventually leads to resolve the dissatisfaction. 
Through this analysis, we can observe the overall flow of how users, while conversing with ChatGPT, experience various dissatisfactions in what proportion, how they respond to them using different tactics, and how this leads to the resolution of these dissatisfactions.
When users encounter dissatisfaction, approximately 34\% opt for \Tno{} while 66\% employ tactics. 
However, it can be seen that approximately 58\% of dissatisfactions are resolved through tactics. 
In the end, users manage to resolve only 28\% of their dissatisfactions using tactics, leaving 72\% of dissatisfactions unresolved.

\subsection{RQ3. Analysis of how dissatisfaction and tactics vary based on the user’s knowledge level of LLMs}
\begin{figure*}[!ht]
    \centering
    \includegraphics[width=0.7\textwidth]{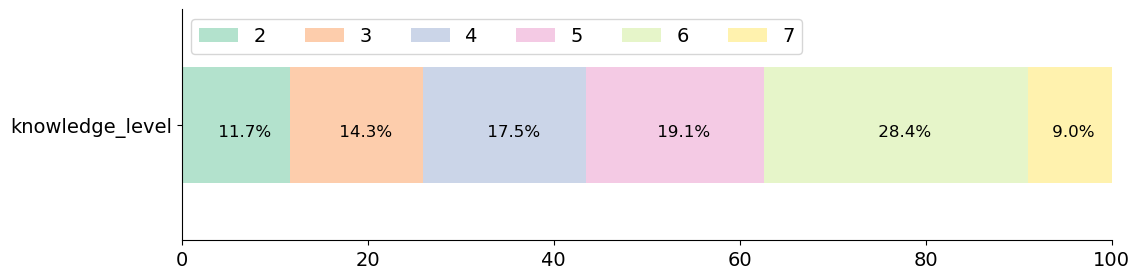}
    \caption{Distribution of participants' knowledge level regarding LLM on a 7-point scale (1: very low, 7: very high). \\
    None of the participants reported a knowledge level of 1.
    }
    \Description{This figure illustrates the distribution of participants' knowledge level regarding LLM on a 7-point scale, where 1 represents knowledge level is very low, and 7 represents knowledge level is very high. In our participants, none of the participants had a knowledge level of 1. 11.7 percent had knowledge level 2, 14.3 percent had knowledge level 3, 17.5 percent had knowledge level 4, 19.1 percent had knowledge level 5, 28.4 percent had knowledge level 6, and 9.0 percent had knowledge level 7.}
    \label{fig:know_level}
\end{figure*}
We analyzed how users’ experience of dissatisfaction and their tactics differ depending on their knowledge levels regarding LLMs.
First, we examined the distribution of users' knowledge levels regarding LLMs in our dataset, as depicted in Fig. ~\ref{fig:know_level}. 
We collected the knowledge level data about LLMs on a 7-point scale, where 1 indicates very low knowledge, and 7 indicates very high knowledge. 
We divided the groups into ``low knowledge level'' (those with a knowledge level 1-3) and `` high knowledge level'' (those with a knowledge level 5-7), as four lies in the middle of the 7-point scale. 
\begin{table*}[!h]
\small{
\centering
\resizebox{0.8\textwidth}{!}{%
\def\arraystretch{1.2}%
\begin{tabular}{c|cc|cc|cc}
\toprule
\multirow{4}{*}{\begin{tabular}[c]{@{}c@{}}Dissatisfaction \\ Category\end{tabular}} & 
    \multicolumn{4}{c|}{Response-level analysis} &
    \multicolumn{2}{c}{User-level analysis} \\ \cline{2-7} &
  \multicolumn{2}{c|}{\textbf{Count: N (\%) *}} &
  \multicolumn{2}{c|}{\makecell[c]{Dissatisfaction \\ Score: mean(std)}} &
  \multicolumn{2}{c}{Frequency: mean (std)} \\ \cline{2-7} 
\multicolumn{1}{c|}{} & high & low & high & low & high & low \\ 
\midrule
\intent{} & 89 (31.56\%) & 45 (29.61\%) & 5.91 (2.85) & 5.18 (3.08) & 0.43 (0.30) & 0.49 (0.39) \\
\depth{} & \textbf{ 50 (17.73\%)} & \textbf{41 (26.97\%)} & 5.02 (2.70) & 5.22 (2.72) & \textbf{0.30 (0.31)} & \textbf{0.38 (0.38)} \\
\acc{} & \textbf{49 (17.38\%)} & \textbf{18 (11.84\%)} & 6.73 (2.85) & 6.5 (2.62) & \textbf{0.24 (0.29)} & \textbf{0.14 (0.21)}\\
\trans{} & 12 (4.26\%) & 9 (5.92\%) & 5.25 (3.33) & 3.67 (3.00) & 0.07 (0.16) & 0.10 ( 0.23)\\
\refuse{} & \textbf{11 (3.90\%)} & \textbf{13 (8.55\%)} & 6.82 (2.79) & 6.92 (2.02) & \textbf{0.07 (0.16)} & \textbf{0.14 (0.26)} \\
\ethic{} & 1 (0.35\%) & 3 (1.97\%) & 3 (-) & 7.33 (2.89) & 0.01 (0.07) & 0.03 (0.08) \\
\format{} & \textbf{70 (24.82\%)} & \textbf{23 (15.13\%)} & 6.66 (2.86) & 5.7 (3.36)  & \textbf{0.28 (0.32)} & \textbf{0.25 (0.37)} \\ 
\bottomrule
\end{tabular}%
}
}
\caption{Dissatisfaction category for knowledge level high and low group (* p-value < 0.01)}
\Description{This table presents the distribution of dissatisfaction categories and corresponding dissatisfaction scores for two groups categorized by knowledge level high and low.}
\label{tab:know-dis}
\end{table*}

\begin{table*}[h]
\centering
\resizebox{\textwidth}{!}{%
\def\arraystretch{1.3}%
\begin{tabular}{cc|cccccccccccc}
\toprule
\multirow{4}{*}{\makecell[c]{Tactic \\ Category}} &
  \multirow{4}{*}{\makecell[c]{Tactic \\ Code}} &
  \multicolumn{8}{c|}{Response-level analysis} &
  \multicolumn{4}{c}{User-level analysis} \\ \cline{3-14} 
 &
   &
  \multicolumn{4}{c|}{Count: N (\%)} &
  \multicolumn{4}{c|}{Effectiveness Score: mean (std)} &
  \multicolumn{4}{c}{Frequency: mean (std)} \\ \cline{3-14}
 &
   &
  \multicolumn{2}{c}{\textbf{Category*}} &
  \multicolumn{2}{c|}{Code} &
  \multicolumn{2}{c}{Category} &
  \multicolumn{2}{c|}{Code} &
  \multicolumn{2}{c}{Category} &
  \multicolumn{2}{c}{Code} \\ \cline{3-14}
 &
   &
  high &
  low &
  high &
  \multicolumn{1}{c|}{low} &
  high &
  low &
  high &
  \multicolumn{1}{c|}{low} &
  high &
  low &
  high &
  low \\ 
\midrule
\multirow{3}{*}{\Trepeat{}} &
  T1 &
  \multirow{3}{*}{16 (6.11\%)} &
  \multirow{3}{*}{\textbf{19 (14.5\%)}} &
  12 (4.4\%) &
  \multicolumn{1}{c|}{9 (6.5\%)} &
  \multirow{3}{*}{\textbf{5.06 (3.00)*}} &
  \multirow{3}{*}{\textbf{2.37 (2.27)*}} &
  4.75 (2.96) &
  \multicolumn{1}{c|}{3.00 (3.00)} &
  \multirow{3}{*}{0.08 (0.17)} &
  \multirow{3}{*}{0.11 (0.25)} &
  0.06 (0.14) &
  0.08 (0.22) \\
 &
  T2 &
   &
   &
  4 (1.5\%) &
  \multicolumn{1}{c|}{12 (8.7\%)} &
   &
   &
  6.00 (3.37) &
  \multicolumn{1}{c|}{1.67 (1.15)} &
   &
   &
  0.02 (0.08) &
  0.04 (0.13) \\
 &
  T3 &
   &
   &
  1 (0.4\%) &
  \multicolumn{1}{c|}{0 (0.0\%)} &
   &
   &
  1 (-) &
  \multicolumn{1}{c|}{- (-)} &
   &
   &
  0.00 (0.02) &
  0.23 (0.35) \\ \hline
\multirow{4}{*}{\Tspecify{}} &
  T4 &
  \multirow{4}{*}{111 (42.37\%)} &
  \multirow{4}{*}{52 (39.7\%)} & 
  84 (30.8\%) &
  \multicolumn{1}{c|}{24 (17.4\%)} &
  \multirow{4}{*}{5.88 (3.56)} &
  \multirow{4}{*}{6 (3.33)} &
  5.77 (3.71) &
  \multicolumn{1}{c|}{7.17 (2.78)} &
  \multirow{4}{*}{0.34 (0.31)} &
  \multirow{4}{*}{0.39 (0.41)} &
  0.23 (0.25) &
  - \\
 &
  T5 &
   &
   &
  13 (4.8\%) &
  \multicolumn{1}{c|}{12 (8.7\%)} &
   &
   &
  5.00 (3.03) &
  \multicolumn{1}{c|}{5.42 (3.73)} &
   &
   &
  0.05 (0.11) &
  0.10 (0.22) \\
 &
  T6 &
   &
   &
  19 (7.0\%) &
  \multicolumn{1}{c|}{13 (9.4\%)} &
   &
   &
  7.00 (2.83) &
  \multicolumn{1}{c|}{6.00 (3.70)} &
   &
   &
  0.09 (0.20) &
  0.07 (0.14) \\
 &
  T7 &
   &
   &
  4 (1.5\%) &
  \multicolumn{1}{c|}{7 (5.1\%)} &
   &
   &
  6.00 (4.08) &
  \multicolumn{1}{c|}{4.00 (2.31)} &
   &
   &
  0.01 (0.05) &
  0.05 (0.18) \\ \hline
\multirow{3}{*}{\Terror{}} &
  T8 &
  \multirow{3}{*}{\textbf{49 (18.70\%)}} &
  \multirow{3}{*}{8 (6.1\%)} &
  37 (13.6\%) &
  \multicolumn{1}{c|}{5 (3.6\%)} &
  \multirow{3}{*}{3.53 (2.60)} &
  \multirow{3}{*}{5.75 (3.06)} &
  3.81 (4.08) &
  \multicolumn{1}{c|}{5.2 (3.83)} &
  \multirow{3}{*}{0.12 (0.24)} &
  \multirow{3}{*}{0.08 (0.18)} &
  0.07 (0.18) &
  0.05 (0.12) \\
 &
  T9 &
   &
   &
  7 (2.6\%) &
  \multicolumn{1}{c|}{2.00 (1.4\%)} &
   &
   &
  3.57 (1.90) &
  \multicolumn{1}{c|}{7.00 (1.41)} &
   &
   &
  0.03 (0.09) &
  0.02 (0.10) \\
 &
  T10 &
   &
   &
  6 (2.2\%) &
  \multicolumn{1}{c|}{2 (1.4\%)} &
   &
   &
  2 (2.45) &
  \multicolumn{1}{c|}{7.00 (1.41)} &
   &
   &
  0.03 (0.12) &
  0.03 (0.12) \\ \hline
\multirow{3}{*}{\Tadapt{}} &
  T11 &
  \multirow{3}{*}{5 (1.91\%)} &
  \multirow{3}{*}{1 (0.8\%)} &
  4 (1.5\%) &
  \multicolumn{1}{c|}{1 (0.7\%)} &
  \multirow{3}{*}{4.40 (3.29)} &
  \multirow{3}{*}{1 (-)} &
  5.25 (3.10) &
  \multicolumn{1}{c|}{1 (-)} &
  \multirow{3}{*}{0.03 (0.11)} &
  \multirow{3}{*}{0.01 (0.07)} &
  0.03 (0.11) &
  0.01 (0.07) \\
 &
  T12 &
   &
   &
  0 (0.0\%) &
  \multicolumn{1}{c|}{0 (0.0\%)} &
   &
   &
  - &
  \multicolumn{1}{c|}{-} &
   &
   &
  - &
  - \\
 &
  T13 &
   &
   &
  1 (0.4\%) &
  \multicolumn{1}{c|}{0 (0.0\%)} &
   &
   &
  1 (-) &
  \multicolumn{1}{c|}{-} &
   &
   &
  0.00 (0.01) &
  - \\ \hline
\multicolumn{2}{c|}{\Tno{}} &
  81 (30.92\%) &
  \textbf{51 (38.9\%)} &
  81 (29.7\%) &
  \multicolumn{1}{c|}{51 (37.0\%)} &
  - &
  - &
  - &
  \multicolumn{1}{c|}{-} &
  0.47 (0.37) &
  0.44 (0.39) &
  0.47 (0.37) &
  0.44 (0.39) \\
\bottomrule
\end{tabular}%
}
\caption{Tactic category and code for knowledge level high and low group (* p-value \textless  0.01)}
\Description{Similar to table 4, this  table shows the analysis results on the count and effectiveness scores on Tactic categories and codes, according to knowledge level high or low groups.}
\label{tab:know-tactic}
\end{table*}
To investigate whether there is a difference in the distribution of dissatisfaction categories between the two groups, we conducted a chi-square test for the dissatisfaction categories of each group and found that there were statistically significant differences in the distribution of dissatisfaction categories by different knowledge groups ($\chi^2$ = 17.7, p-value < 0.01). 
Specifically, we observed that the low-knowledge group experiences \depth{} ((count: 26.97\%, user-level frequency: 0.38)) and \refuse{} (count: 8.55\%, user-level frequency: 0.14) more frequently, while the high-knowledge group experiences \acc{} (count: 17.38\%, user-level frequency: 0.24) and \format{} (count: 24.82\%, user-level frequency: 0.28) more frequently.
On the other hand, we conducted a Mann-Whitney U test to investigate the differences in dissatisfaction scores between knowledge groups, but there were no significant differences.

\begin{figure*}[h]
    \centering
    \subfigure[]{
        \includegraphics[width=0.485\textwidth]{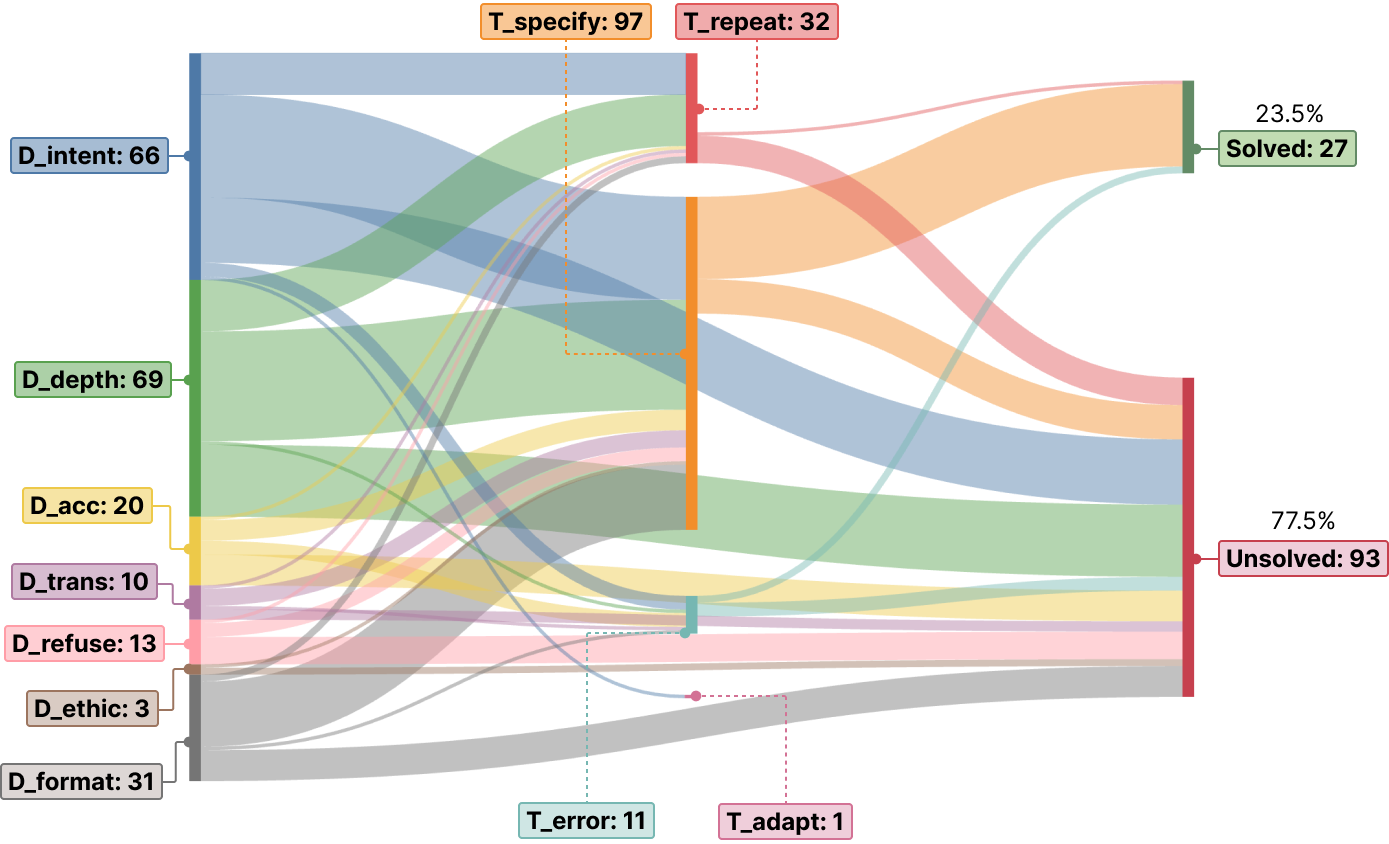}
        \label{fig:sankey_low}
    }
    \subfigure[]{
        \includegraphics[width=0.485\textwidth]{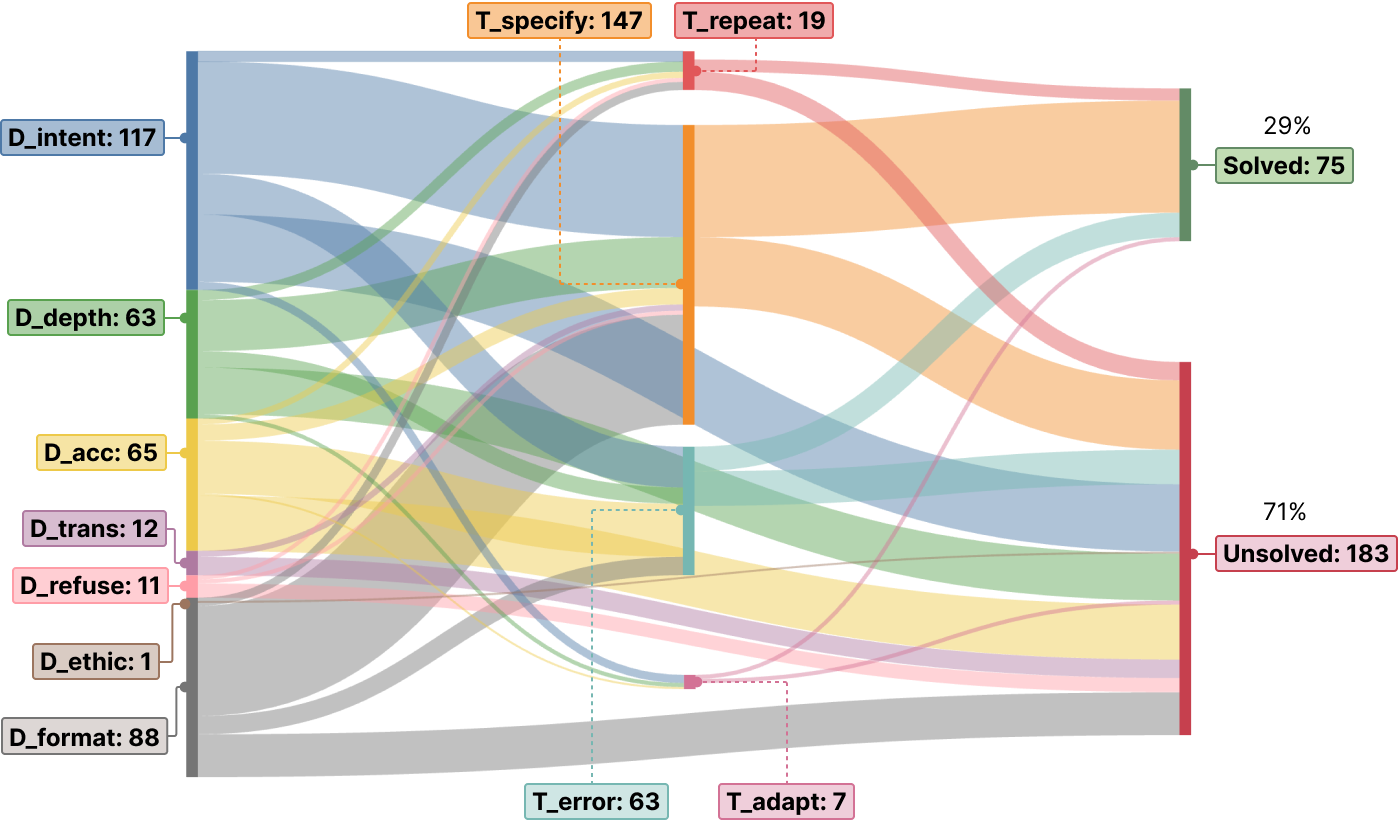}
        \label{fig:sankey_high}
    }
\caption{
Sankey diagrams by users' knowledge level of LLMs that visualize how users respond among four tactic categories after experiencing each of the dissatisfaction categories and finally whether that dissatisfaction was solved or not. (a): Low-knowledge group's Sankey diagram (b): High-knowledge group's Sankey diagram.
}
\Description{There are two figures, each of them is a Sankey diagram by users' knowledge level of LLMs that visualizes how users respond to each of the dissatisfaction categories using which tactic, and finally whether it was solved or not. (a): The low-knowledge group's Sankey diagram (b): The high-knowledge group's Sankey diagram}
\end{figure*}

Similarly, we conducted a chi-square test for tactic categories and found significant differences in the count of tactic categories among the two groups ($\chi^2$ = 21.6, p-value < 0.01).
In particular, \Tno{} was more prevalent in the low-knowledge group.
Additionally, \Trepeat{}, which involves minimal prompt engineering, was more commonly used in the low-knowledge group, while \Terror{}, aimed at pointing out and rectifying errors in ChatGPT responses, was more prevalent in the high-knowledge group.
Furthermore, to compare and analyze the effectiveness of the tactics used in each knowledge group, we performed a Mann-Whitney U test on the effectiveness scores of tactic categories, which were collected from users. 
Through this test, we found that the effectiveness of the \Trepeat{} was statistically higher in the high-knowledge group (p-value < 0.01, effect size= 0.5789).
Fig.~\ref{fig:sankey_low} and \ref{fig:sankey_high} present Sankey diagrams that illustrate how users in the low-knowledge and high-knowledge groups experience dissatisfaction categories from \nobreak ChatGPT's responses, respond to the dissatisfactions with each tactic category at user prompts, and whether these tactics ultimately resolve their dissatisfactions or not. Through this, we can see that the rate of resolving dissatisfaction in the high-knowledge group (29\%) is higher than low-knowledge group (23.5\%).

\section{Discussion}
In this section, we first discuss the interpretation of our results and their implications. Second, we suggest design implications for building LLMs with better usability based on our study results. Lastly, we discuss the limitations of our study and future work.

\subsection{Interpretation of results}
Building upon the analysis of user-side dissatisfaction and corresponding user tactics during the conversation, we discuss the most prevalent, severe, and unaddressed categories of dissatisfaction and their implications.
We also discuss the differences in dissatisfaction and corresponding tactics across users with different knowledge levels about LLM.

\subsubsection{The Most Prevalent Dissatisfaction and Tactics.} \label{sec:discussion_intent}
Our results suggest that despite the advances in LLMs to align with the user intent, there still exists much room for improvement from the users' perspective.
With recent advancements in LLMs and the introduction of techniques to align LLMs with user intents, such as Reinforcement Learning from Human Feedback (RLHF), LLMs are now known to better align with human intent than before ~\cite{ouyang2022training, christiano2017deep, ziegler2019fine}.
However, we found that \intent{}, the dissatisfaction in terms of understanding users' intent, is the most prevalent (Table~\ref{tab:dis}) and frequently co-occurring with other dissatisfaction categories (Figure~\ref{fig:co_occr_nor}).
We also discovered that users frequently use \Tspecify{} that further specify their intent to address the dissatisfaction. Moreover, users rated \Tspecify{} as the most effective among tactic categories, but there are still many cases (about 42\%) where dissatisfaction was not resolved despite using this tactic. 
This may be because users have difficulty clearly representing their intent. 
Previous work on web search and information retrieval has also noticed this problem~\cite{10.1145/1526709.1526773}, and there exist several methods to better support users to specify their intent in these domains, such as context-sensitive query auto-completion~\cite{10.1145/1963405.1963424} and context-based term suggestions~\cite{RIEH2006751}.
Similarly, in the context of LLMs, further research can investigate methods to help users specify their intents based on their context.

\subsubsection{The Most Severe or Unaddressed Dissatisfaction}
Self-reported scores on the level of dissatisfaction show that users perceived the dissatisfaction of \acc{} to be the most severe (Table~\ref{tab:know-dis}). 
This shows that users feel a high level of dissatisfaction with limitations of LLMs related to information accuracy such as hallucination ~\cite{Ye2023CognitiveMA, bang2023multitask, ji2023survey, liu2023trustworthy}, inconsistency or incorrectness in the responses ~\cite{Jang2022BECELBF, elazar2021measuring, liu2023trustworthy, jang2023consistency}, and the inability of ChatGPT to provide up-to-date information ~\cite{zhao2023survey, alkhamissi2022review}.
Furthermore, our findings show that users tend to respond to this dissatisfaction primarily by pointing out LLM's faults or correcting them (\Terror{}), but more than half of them (57\%) nevertheless fail to resolve this dissatisfaction.

We also found that when users encountered dissatisfaction when their prompts were refused to answer (\refuse{}), when ethical concerns or biases were found in the response (\ethic{}), or when they had a lack of understanding of the internal logic of the generated response (\trans{}), they often did not attempt to address the dissatisfaction or even terminated the conversation.
For instance, one user explained their decision to end the conversation as follows: ``I ended the conversation as I felt like there was no common understanding and was not looking forward to explaining myself any further than my original prompt.'' 
Through this, we can see that if the users experience such dissatisfaction, they not only have difficulty communicating with ChatGPT but also have no idea how to further improve their prompts, often terminating the conversation.

One notable point here is that \ethic{} and \refuse{} can be in a trade-off relationship.
Including OpenAI~\footnote{https://openai.com/}, the company that developed ChatGPT, many companies have adopted a strategy where the LLM avoids answering when faced with potentially unethical or biased prompts, responding with statements like ``As a language model, I am not capable of ... ''~\cite{Zheng2023WhyDC, borji2023categorical}.
Although companies could avoid being embroiled in ethical issues, this approach might have introduced another dimension of dissatisfaction (\refuse{}) for users. 
Self-reported scores on the level of dissatisfaction show the level of severity for both \ethic{} and \refuse{} are similar (Table~\ref{tab:know-dis}). This suggests that the current approach of refusing to answer instead of giving responses with ethical concerns may not reduce users' overall dissatisfaction. Thus, it is necessary to find other measures that could also lower the users' dissatisfaction when faced with unethical or biased prompts. 

\subsubsection{Differences in Dissatisfaction and Corresponding Tactics Across LLM Knowledge Levels}
Our result revealed that there exist significant differences in dissatisfaction and employed tactics between high- and low-knowledge user groups.
We observed that the low-knowledge group reports higher occurrences of \depth{}---dissatisfaction that ChatGPT's response is too general and lacks detail or originality---than the high-knowledge group (Table~\ref{tab:know-dis}). 
One possible reason behind this is that the low-knowledge user group might have overestimated ChatGPT's creative capabilities. This could be because low-knowledge user groups may be more prone to unconditionally accepting media or news which states that ChatGPT can perform creative tasks such as writing poetry and song lyrics ~\cite{dwivedi2023so, chatGPTi85online}. This may have led them to expect more creative responses, resulting in a higher possibility of feeling disappointment.
In contrast, the high-knowledge group may have possessed a better understanding of ChatGPT's limitations. Knowing that ChatGPT's responses are based on trained patterns from existing datasets could have allowed them to be more generous towards the responses that lack originality.
We speculate that the low-knowledge group might have a less accurate mental model of the capacity of LLM, misunderstand its capabilities, and experience more dissatisfaction in terms of \depth{}.

Moreover, the tactics employed in response to these dissatisfactions differed between the two groups. 
Compared to the high-knowledge group, the low-knowledge group relied more on `No tactic' and more frequently used \Trepeat{}, which requires minimal effort for prompt writing (Table~\ref{tab:know-tactic}). 
This may be because the low-knowledge users may not know much about the various options of tactics they could take. 
Interestingly, however, although high-knowledge users used \Trepeat{} less, they found it more effective in solving their dissatisfaction. This may indicate that high-knowledge users tend to have a better sense of when is the right time to use \Trepeat{}.

\subsection{Design Implications for Building LLMs with Better Usability}
Based on our study result, we suggest three design implications to enhance the usability of LLMs: (1) supporting users to represent their intent, (2) recommending effective multi-turn prompt tactics to users, and (3) providing personalized LLM experiences to users. 
\subsubsection{Supporting users to represent their intent better}
We suggest a design that facilitates a better representation of the user's intent. 
In the current system interface, there is a lack of design support to help users' prompt writing process, and we found that users frequently face limitations in conveying their full intent in Sec ~\ref{sec:discussion_intent}. 
To address these challenges and facilitate a better representation of the user's intent, it is necessary to have a design that helps users refine their prompts to align them more precisely with their intent. 
This design could involve tokenizing user prompts and using this as a basis to offer keyword-specific suggestions. For example, if a user writes a prompt, ``Explain recent issues related to Autonomous Vehicles (AVs) in simple terms.'', keywords can tokenize the prompt, and the following keyword-specific suggestions can be provided: the types of AVs, the time frame for recent, the types of issues (e.g., ethical), and the appropriated level of simplicity for the terms used. Moreover, considering dissatisfaction arising from extensive and detailed responses (\format{}), giving suggestions utilizing multi-modality, such as image and video, could enable a better user experience when they can succinctly represent the users' intent.
This allows users to refine their prompts by selecting the suggestions, ensuring a more accurate alignment with their intent. Providing users with a range of suggestions and enabling them to select suggestions by reflecting their intent can empower users to express their intent effectively. 

\subsubsection{Recommending effective multi-turn prompt tactics to users}
To enhance user satisfaction during multi-turn interactions with LLM, we suggest a design that recommends effective prompt tactics to users during the conversation. Our public dataset could be utilized for this process since it contains various prompt tactics (Table ~\ref{tab:tactic-result}) and their effectiveness reported by users to address their dissatisfaction.
For instance, an interaction can be envisioned where the LLM predicts the probability of user dissatisfaction with a generated response. If the probability is high, the system can proactively guide users to employ some effective tactics in their subsequent prompt to address the anticipated dissatisfaction. 

We also recommend evolving this design to incorporate effective prompt engineering techniques suitable for multi-turn interactions, such as Chain-of-Thought (CoT)~\cite{zhang2022automatic}. While a thread of research has addressed effective prompt engineering techniques to get desired responses from LLMs, they usually focus on crafting one prompt. Moreover, there is a lack of research on prompt engineering techniques tailored to address or mitigate user dissatisfaction during conversations.
By integrating our data-driven insights on users' effective prompt tactics with prompt engineering techniques, we propose that recommending tactics to users during multi-turn interactions will yield more favorable responses, enhancing their overall satisfaction.

\subsubsection{Providing personalized LLM experience}
We suggest the need for a design that provides personalized LLM experiences based on our finding that there exist differences in dissatisfaction and corresponding tactics depending on the user's level of knowledge about LLMs.
One of the possible designs for personalized LLM experiences is to adjust the refusal policies or attitudes that LLM refuses to answer according to the user's knowledge levels about LLM. 
This is because our results show that the low-knowledge group experienced more dissatisfaction with ChatGPT's refusal to answer (\refuse{}) than the high-knowledge group. This may be because the low-knowledge group tends to ask more questions that were limited for ChatGPT to answer without fully understanding ChatGPT's capabilities.
Thus, rather than responding with a generic ``As a language model, I am not capable of...'' a more direct explanation addressing its limitations to better inform users of its capability may be required for low-knowledge users. 

To facilitate personalized LLM experiences, we emphasize the need for user modeling based on prior sessions where LLM can gain information about the user's state before chatting. The user's state encompasses not only their knowledge level about LLM but also their usage purpose, specific task at hand, the language or proficiency level they used for chatting, and more. Such sessions serve to shape the user's mental model of LLM and vice versa, fostering a mutual understanding. Through this approach, users can benefit from customized interactions that consider their individual circumstances, ultimately improving their overall LLM experience.

\subsection{Generalization of Results}
While our study utilized ChatGPT as a case study, our research methodology and its implications can extend beyond ChatGPT. 
The seven categories of user-side dissatisfaction identified through our SLR (Table ~\ref{tab:slr-result}) encompassed references that span various LLMs.
Hence, leveraging these categories and our analysis method, future research can apply similar investigations to different LLMs.
In addition, the four categories of user tactics (Table ~\ref{tab:tactic-result}) were derived from analyzing user behavior patterns in multi-turn conversations with LLM, based on ChatGPT user data. 
The consistent nature of user behavior across various LLMs with similar multi-turn chat-based interfaces suggests potential generalizability to other LLMs.

However, it is essential to consider potential variations that may arise due to specific features in LLMs, technical advancements, and changes in user perception towards LLMs.
Even if the categories remain constant, their distribution and severity may change.
For instance, we can expect that while the proportion of dissatisfaction arising from accuracy (\acc{}) might decrease as the performance of LLM improves, its perceived severity may intensify as user's expectations towards LLMs get higher.
Furthermore, while many users may currently lack awareness of the ethical issues related to LLM, \ethic{} might increase as they become informed about the potential ethical threats posed by LLM.
Therefore, extending our analysis to different LLMs or the same LLM over time allows for a comprehensive comparison of user dissatisfaction across various models and versions, providing insights into the direction of evolving LLMs.
Our data analysis also showed that user dissatisfaction varied based on users' knowledge level regarding LLMs. From this, we may refer to the dissatisfaction distribution of the current high-knowledge group while inferring the dissatisfaction distribution of LLM users in the future.
In terms of user tactics, the emergence of novel interaction components beyond chat-based interfaces may lead to different user behavior patterns, which would require further investigation.

\subsection{Limitations and Future Work}
We present the limitations of our work and possible future work.

First, our analysis is based on self-reported data from users. We tried to ensure the quality of the data by careful filtering and pre-processing of the data while checking on the actual conversation log. However, dissatisfaction levels and tactic effectiveness are based on participants' self-reported scores, which may suffer from subjectiveness and heavily rely on the participant's memory. We also tried to eliminate this problem by only collecting conversation logs within 1 month, but the problem may still linger.

Second, we investigated the difference in user dissatisfaction and tactics according to the difference in knowledge level of LLMs. Future work can expand on our work and further examine whether the differences in dissatisfaction and tactics exist according to other dimensions. For instance, since LLMs are chat-based, there may exist differences between those different English proficiency. Moreover, since users may have different expectations according to tasks, there may exist differences when given different tasks. For instance, fact-oriented tasks, such as finding information or explaining a real-world fact, will have more relevance with \acc{} since the user expects to get correct information. On the other hand, creative tasks, such as writing stories or scenarios, will have less relevance with \acc{} but more relevance with \intent{}, since users will be interested in how well the LLM can understand their needed content or context of creating content to their situations.

Lastly, our analysis of user-side dissatisfaction and tactics was based on ChatGPT user data. 
Therefore, there may be some differences in how users undergo dissatisfaction depending on other LLMs. For instance, specific wordings used when LLM refuses to answer can affect how much users feel dissatisfaction regarding \refuse{}. Moreover, since \acc{} is a category that is directly related to the performance of LLMs, users may face different levels of dissatisfaction for \acc{}.
\section{Conclusion}
In this study, with ChatGPT as the case study, we explored user-side dissatisfaction and corresponding tactics during the conversation with chat-based LLM. 
Through a systematic literature review, we identified seven categories of user-side dissatisfaction from LLM-generated responses. 
Then, we collected data from 107 users conversing with ChatGPT, and uncovered prevalent, severe, and unaddressed dissatisfactions.
We also analyzed four users' tactic categories to address their dissatisfaction and their prevalence and effectiveness. 
We also investigated how these vary depending on the users' knowledge level of LLMs.
Our findings provide insights into how LLM and its interface can be further developed to aid people when they encounter dissatisfaction.
One potential is user-side prompt engineering techniques that can be utilized in the middle of the conversation when dissatisfaction occurs. The pair of dissatisfactions and corresponding tactics can guide this prompt engineering.
In addition to these contributions, we have made a publicly accessible dataset available, containing actual user conversation data related to dissatisfaction. 
This research deepens the understanding of user dissatisfaction in LLM interactions, providing a foundational knowledge base for future enhancements that can benefit users across knowledge levels.
\begin{acks}
This work was supported by Institute of Information \& Communications Technology Planning \& Evaluation (IITP) grant funded by the Korea government(MSIT) (No.2019-0-00075, Artificial Intelligence Graduate School Program (KAIST)).
This work was also supported by Institute of Information \& Communications Technology Planning \& Evaluation (IITP) grant funded by the Korea government (MSIT) (No.2021-0-01347, Video Interaction Technologies Using Object-Oriented Video Modeling).
\end{acks} 

\bibliographystyle{ACM-Reference-Format}
\bibliography{references}
\appendix
\section{Appendix}

\subsection{Systematic Literature Review Paper List}\label{appendix:slr_all}
All paper lists corresponding user-side dissatisfaction codes are in Table ~\ref{tab:slr_full}.
\begin{table*}[!ht]
\scriptsize{
\begin{center}
\resizebox{\textwidth}{!}{%
\def\arraystretch{1.6}%
\begin{tabular}{p{0.17\columnwidth}p{0.56\columnwidth}l}
\toprule
    \textbf{Category (7)} & \textbf{Code (19)} & \textbf{Literatures} \\
\midrule
    \multirow{3}{0.15\columnwidth}{\makecell[l]{Intent \\ Understanding\\ \textbf{(\intent{})}}}  & C1. Response does not meet users' intent or instruction. & \cite{kaddour2023challenges,Dong_2022,borji2023categorical,si2023measuring,kumar2023analysis,rao2023evaluating} \\ \cline{2-3}
& C2. Response is not aligned with the user's context. &  \cite{rao2023evaluating,borji2023categorical,sallam2023chatgpt,bubeck2023sparks,
si2023measuring,guo2023close,kumar2023analysis,khan2023chatgpt,weidinger2021ethical,ray2023chatgpt,farrokhnia2023swot,
kaddour2023challenges,brown2022does,Dong_2022} \\ \cline{2-3}
 & C17. The tone or communication style is disappointing. & \cite{guo2023close,borji2023categorical,sallam2023chatgpt,azaria:hal-03913837,ray2023chatgpt} \\ 
\hline{}
 
\multirow{3}{0.15\columnwidth}{\makecell[l]{Content Depth \\ and Originality \\ \textbf{(\depth{})}}}  & C3. Response is too general. & \cite{borji2023categorical,zhao2023survey,guo2023close,kumar2023analysis,sharma2023chatgpt,yeo2023assessing,ray2023chatgpt} \\ 
\cline{2-3} 
 & C4. Response lacks originality. & \cite{kitamura2023chatgpt,borji2023categorical,sallam2023chatgpt,bubeck2023sparks,guo2023close,thorp2023chatgpt,ray2023chatgpt,lecler2023revolutionizing,dwivedi2023so} \\ 
 \cline{2-3} 
 & C5. Response lacks information. & \cite{kumar2023analysis,kasneci2023chatgpt,borji2023categorical,bubeck2023sparks,
 guo2023close,mann2023artificial,rao2023evaluating,sharma2023chatgpt,
 yeo2023assessing,nisar2023chatgpt,ray2023chatgpt,
 lecler2023revolutionizing,farrokhnia2023swot,Dong_2022}\\ 
 \hline{}

\multirow{6}{0.15\columnwidth}{\makecell[l]{Information \\ Accuracy \\ \textbf{(\acc{})}}}  & C6. The response contains incorrect information. & \makecell[l]{\cite{azaria:hal-03913837, kasneci2023chatgpt,liu2023trustworthy,cao2023comprehensive,borji2023categorical,qadir2023engineering,sallam2023chatgpt,floridi2023ai,zhang2023small,elazar2021measuring,bubeck2023sparks,alkhamissi2022review,zhou2023navigating,jin2023can,kumar2023analysis} \\ \cite{thorp2023chatgpt,wang2023can,rao2023evaluating,sharma2023chatgpt,duong2023analysis,min2023recent,weidinger2022taxonomy,weidinger2021ethical,ray2023chatgpt,lecler2023revolutionizing,farrokhnia2023swot,dwivedi2023so,gill2023chatgpt,hadi2023survey,gpt4syst42:online} }\\ 
\cline{2-3} 
& C7. Response is based on training data cut off at a certain date, and has limited access to newly created data. & \cite{yeo2023assessing,thirunavukarasu2023large,zhao2023survey,sallam2023chatgpt,alkhamissi2022review,guo2023close,khan2023chatgpt,farrokhnia2023swot,kaddour2023challenges,dwivedi2023so,Dong_2022} \\ 
 \cline{2-3}
 & C8. Response is inconsistent. & \cite{alkhamissi2022review,liu2023trustworthy,zhao2023survey,jang2023consistency,elazar2021measuring,bubeck2023sparks,liu2023evaluating,zhou2023navigating,holzinger2023ai,wang2023can,duong2023analysis,gill2023chatgpt} \\ 
 \cline{2-3}
 & C9. ChatGPT struggles with reasoning. & \cite{zhang2023small,liu2023trustworthy,zhao2023survey,borji2023categorical,bang2023multitask,qin2023chatgpt,alkhamissi2022review,jin2023can,mann2023artificial,khan2023chatgpt,sharma2023chatgpt,farrokhnia2023swot,dwivedi2023so,hadi2023survey} \\ 
 \cline{2-3}
 & C10. (Hallucination) ChatGPT fabricates contents that conflict with the source content or cannot be verified from existing sources. & \cite{ji2023survey,liu2023trustworthy,thirunavukarasu2023large,zhao2023survey,cao2023comprehensive,borji2023categorical,bang2023multitask,weidinger2022taxonomy,tian2023opportunities,kaddour2023challenges,gill2023chatgpt,hadi2023survey,gpt4syst42:online} \\ 
 \cline{2-3} 
 & C19. (Sycophancy) ChatGPT excessively conforms to the user. & \cite{bang2023multitask,liu2023trustworthy,guo2023close,perez2022discovering} \\ 
\hline{}

Transparency \textbf{(\trans{})} & C11. It's difficult to understand the reasons, criteria, logic, and evidence behind the responses. & \cite{thirunavukarasu2023large,liu2023trustworthy,sallam2023chatgpt,bang2023multitask,jang2023consistency,bubeck2023sparks,alkhamissi2022review,wang2023can,rao2023evaluating,sharma2023chatgpt,ray2023chatgpt,dwivedi2023so,gill2023chatgpt,hadi2023survey} \\ 
\hline{}

\multirow{3}{0.15\columnwidth}{\makecell[l]{Refusing to \\ Answer \\ \textbf{(\refuse{})}}}  & C12. ChatGPT avoids giving its own opinion by saying something similar to ``As a language model, I am not capable …'' & \cite{borji2023categorical,guo2023close} \\ 
\cline{2-3} 
 &  C13. ChatGPT avoids talking about difficult or controversial issues by saying something similar to ``As a language model, I am not capable ...'' & \cite{bang2023multitask,guo2023close} \\ 
\cline{2-3}
 &  C7. Response is based on training data cut off at a certain date, and has limited access to newly created data. & \cite{yeo2023assessing,thirunavukarasu2023large,zhao2023survey,sallam2023chatgpt,alkhamissi2022review,guo2023close,khan2023chatgpt,farrokhnia2023swot,kaddour2023challenges,dwivedi2023so,Dong_2022} \\
\hline{}
 
\multirow{3}{0.15\columnwidth}{\makecell[l]{Content Ethics \\ and Integrity \\ \textbf{(\ethic{})}}} &  C14. Response contains unlawful content & \cite{liu2023trustworthy,holzinger2023ai} \\ 
\cline{2-3}
&  C15. Response contains unethical, harmful content. & \cite{weidinger2021ethical,liu2023trustworthy,thirunavukarasu2023large,sallam2023chatgpt,bender2021dangers,holzinger2023ai,sharma2023chatgpt,min2023recent,weidinger2022taxonomy,gehman2020realtoxicityprompts,weidinger2021ethical,brown2022does,gpt4syst42:online} \\ 
\cline{2-3} 
&  C16. Response contains biased content. & \cite{cao2023comprehensive, kasneci2023chatgpt,liu2023trustworthy,thirunavukarasu2023large,borji2023categorical,sallam2023chatgpt,antaki2023evaluating,sharma2023chatgpt,min2023recent,abid2021persistent,wolfe2022american,gadiraju2023wouldn,weidinger2022taxonomy,lucy2021gender,navigli2023biases,ray2023chatgpt,farrokhnia2023swot,gill2023chatgpt,hadi2023survey}  \\
\hline{}

\multirow{3}{0.15\columnwidth}{\makecell[l]{Response \\ Format and \\ Attitude \\ \textbf{(\format{})}}}  & C17. The tone or communication style is disappointing. & \cite{guo2023close,borji2023categorical,sallam2023chatgpt,azaria:hal-03913837,ray2023chatgpt} \\ 
\cline{2-3} 
&  C18. Response is overly detailed or too long & \cite{yang2023exploring,sallam2023chatgpt,bubeck2023sparks,guo2023close,rao2023evaluating} \\ 
\cline{2-3} 
&  C19. (Sycophancy) ChatGPT excessively conforms to the user. & \cite{bang2023multitask,liu2023trustworthy,guo2023close,perez2022discovering} \\ 
\bottomrule
\end{tabular}
}
\end{center}
}
\caption{7 category and corresponding 19 codes of user-side dissatisfaction from LLM Responses.}
\Description{Similar to Table one, this table summarizes the result of systematic literature review on user dissatisfaction. There are four columns: Category(7), Description, Code(19), and Example. Each dissatisfaction category has one description and one or more specific dissatisfaction codes. Examples are research papers related to each code, but we list all research papers rather than just one as we did in Table one.}
\label{tab:slr_full}

\end{table*}

\subsection{Data Filtering Criteria and Detailed Reason}\label{appendix:criteria}
\subsubsection{Conversation-Level Filtering}

\smallskip
\noindent
\textbf{\textit{Conversation older than 30 days.}}
We collected real-world experience data from individuals, which inherently consists of past data they have encountered. \nobreak Therefore, in order to encourage respondents to recall these past experiences while responding to our data collection system, we restricted the chat dates to ``previous 30 days'' from the survey date. 
Although the survey included explicit instructions regarding this matter, we identified four cases where participants reported chat dates older than 30 days, and excluded them. 

\smallskip
\noindent
\textbf{\textit{Conversation with a memory level of 3 or lower.}}
Even if a conversation occurred within the previous 30 days, it was considered unreliable if the user had a low memory level regarding the conversation. 
Therefore, conversations where the user's memory level was rated 3 or lower on a 7-point scale were filtered out. This criterion led to the exclusion of five conversations.

\smallskip
\noindent
\textbf{\textit{Conversation for fun or testing purposes.}}
Our research focused on real-world experiences related to dissatisfaction encountered while using LLMs for practical purposes.
Therefore, we do not delve into scenarios where users intentionally provoke dissatisfactory responses from LLMs, attempting to manipulate the model's behavior through techniques like jailbreaking~\cite{LLMJailb71:online, liu2023jailbreaking}, using LLM solely for fun or testing.
Despite the explicit instructions regarding this in the data collection system, seven conversations were identified as falling into this category and were filtered out.

\smallskip
\noindent
\textbf{\textit{Conversation from versions other than GPT-3.5.}}
Considering the significant differences in performance between GPT-3.5 and GPT-4 ~\cite{gpt4syst42:online}, we also considered the GPT version used in the conversation. Four conversations used GPT-4, while all others used GPT-3.5. To maintain data consistency, we filtered out the four conversations that used GPT-4.

\subsubsection{Response-Level Filtering}

\smallskip
\noindent
\textbf{\textit{Dissatisfaction due to ChatGPT's error messages}}
Dissatisfaction caused by ChatGPT responses being interrupted or encountering errors was not our research scope. Three responses fell under this category. 

\smallskip
\noindent
\textbf{\textit{Unconvincing dissatisfaction}}
Seven cases were identified where it was challenging to understand why the user was dissatisfied when reviewing both the ChatGPT conversation and the user's dissatisfaction reasons. 

\smallskip
\noindent
\textbf{\textit{Mismatch between score and reason}}
In one case, the effectiveness score for resolving dissatisfaction was 1 (indicating not effective), but the reason for that score was reported that the dissatisfaction was resolved by the prompt. This mismatch led to the exclusion of this case. 

\smallskip
\noindent
\textbf{\textit{No Correlation between selected dissatisfactions and subsequent prompts for resolving that dissatisfaction}}
In five cases, we observed a lack of correlation between selected dissatisfaction categories and selected subsequent prompts to address such dissatisfaction. For example, it was the case a prompt that had nothing to do with the selected dissatisfaction was chosen to resolve the dissatisfaction.





\end{document}